\documentclass[conference,9pt]{IEEEtran}
\usepackage{ifthen}
\usepackage[utf8]{inputenc}
\usepackage{latexsym, dsfont}
\usepackage{amsmath,amsthm,amssymb}
\usepackage[inline]{enumitem}
\usepackage{comment}
\usepackage{siunitx}
\usepackage{graphicx}
\usepackage{caption}
\usepackage[capitalize]{cleveref}
\usepackage{todonotes}
\usepackage[caption=false,font=footnotesize]{subfig}
\usepackage{mathtools}
\usepackage{tikz}
\newboolean{PGFPlots}
\setboolean{PGFPlots}{true}
\usepackage[inline]{enumitem}  
\makeatletter
\newcommand{\inlineitem}[1][]{%
\ifnum\enit@type=\tw@
    {\descriptionlabel{#1}}
  \hspace{\labelsep}%
\else
  \ifnum\enit@type=\z@
       \refstepcounter{\@listctr}\fi
    \quad\@itemlabel\hspace{\labelsep}%
\fi}

\ifthenelse{\boolean{PGFPlots}}{
\usepackage{pgfplots}
\usepgfplotslibrary{units,groupplots}
\usetikzlibrary{external}
\tikzexternalize
\tikzset{external/only named=true}
}{}

\newcommand{\dupD}{\delta^\uparrow}
\newcommand{\ddoD}{\delta^\downarrow}

\newcommand{\dsu}{\delta_S^\uparrow}
\newcommand{\dsd}{\delta_S^\downarrow}

\newcommand{\figPath}[1]{figures/#1}

\newcommand{\NOR}{\texttt{NOR}}
\newcommand{\NAND}{\texttt{NAND}}

\newcommand{\AOI}{\texttt{AOI}}
\newcommand{\C}{\texttt{C}}

\newcommand{\dmin}{\delta_{\mathrm{min}}}

\newcommand{\spice}{\textit{SPICE}}

\newcommand{\vth}{V_{th}}

\newcommand{\vdd}{V_{DD}}
\newcommand{\gnd}{\textit{GND}}
\newcommand{\vout}{V_{out}}

\newcommand{\dd}{\mathrm{d}}

\newcommand{\nmos}{nMOS}
\newcommand{\pmos}{pMOS}

\newcommand{\ohm}{(OHM)}

\newcommand{\on}{\mbox{\emph{on}}}
\newcommand{\off}{\mbox{\emph{off}}}

\newboolean{conference}
\setboolean{conference}{false}  

\crefname{equation}{}{}

\newtheorem{theorem}{Theorem}

\begin{document}

\ifthenelse{\boolean{conference}}
{
\title{Accurate Hybrid Delay Models for Dynamic Timing Analysis}
}
{
\title{An Accurate Hybrid Delay Model for Multi-Input Gates}
}

\ifthenelse{\boolean{conference}}
{}
{
\author{
  \IEEEauthorblockN{
    Arman Ferdowsi,
    Ulrich Schmid,
    and Josef Salzmann
  }
  \IEEEauthorblockA{Embedded Computing Systems Group (E191-02)\\
    TU Wien, Vienna, Austria \\
    \{aferdowsi, s, jsalzmann\}{@}ecs.tuwien.ac.at}
}
}

\maketitle

\begin{abstract}
In order to facilitate the analysis of timing relations between individual transitions in a signal trace,
dynamic digital timing analysis offers a less accurate but much faster alternative to analog simulations 
of digital circuits. This primarily requires gate delay models that also account for the fact that the 
input-to-output delay of a particular input transition also depends on the temporal distance to the previous 
output transitions.
In the case of multi-input gates, the delay also experiences variations caused by multi-input switching (MIS) effects, i.e., transitions at different inputs that occur in close temporal proximity. In this paper, we advocate the development 
of hybrid delay models for CMOS gates obtained by replacing transistors with time-variant resistors. We exemplify
our approach by applying it to a NOR gate (and, hence, to the dual NAND gate) and a Muller \C\ gate.
We analytically solve the resulting first-order differential equations with non-constant-coefficients, and derive analytic expressions for the resulting MIS gate delays. The resulting formulas not only pave the way to a sound model parametrization procedure, but are also instrumental for implementing fast and efficient digital timing simulation. By comparison with analog simulation data, we show that our models faithfully represent all relevant MIS effects. Using an implementation in the Involution Tool, we demonstrate that our model surpasses the alternative digital delay models for NOR gates known to us in terms of accuracy, with comparably short running times.
\end{abstract}

	
	\section{Introduction}
	\label{sec:intro}

%
%
%

\emph{Digital} timing analysis techniques are essential for modern circuit 
design. Thanks to the elaborate static timing analysis techniques available
today, which employ detailed models like CCSM~\cite{Syn:CCSM} and 
ECSM~\cite{Cad:ECSM} that facilitate a very accurate corner case analysis 
of the delays of the gates making up a digital circuit, essential 
circuit timing properties like worst-case critical path delays can nowadays be 
determined very accurately and very fast.

Still, asynchronous digital circuits may contain parts where corner-case delay
estimates are not sufficient for understanding the overall behavior and for 
validating correct operation. We exemplify
this by means of the token-passing ring described and analyzed by Winstanley, Garivier and Greenstreet 
in \cite{CharlieEffect}:\footnote{\ifthenelse{\boolean{conference}}
{
A more recent application example would be delay-encoded 
inter-neuron links in hardware-implemented spiking neural networks~\cite{BVMRRVB19},
where the delay between successive transitions must be tracked when traveling over the link
for verifying correct operation.}
{
A more recent application example would be \emph{spiking neural network} (SNN) hardware
implementations \cite{Fur16, BVMRRVB19},
where the delay between successive transitions must be tracked when traveling over the link
for verifying correct operation: Neurons in
SNNs communicate via discrete-value continuous-time signals, where
analog values are effectively encoded via the delay between successive
spike events. Besides rate-based approaches, where a single analog
value is represented by the frequency of a whole spike train,
\emph{inter-spike interval} (ISI) time encoding uses the time
interval between two spikes to encode this information in a more
energy-efficient way. Obviously, for the latter, it is crucial that
the actual implementation of such a communication link preserves
delays between successive signal transitions as accurately as possible.
}} 
Stages consisting of a 2-input Muller \C\ gate, with its inputs connected to the 
previous resp.\ next stage, implement a self-timed ring oscillator. A detailed analysis
revealed that it exhibits two different
modes of operation, namely, burst behavior versus evenly spaced output transitions, 
with unpredictable mode switches between those. The actual operation mode depends on the 
subtle interplay between two effects that determine the delay
of a Muller \C gate: the \emph{drafting effect}, a decrease of 
the delay happening when the resulting output transition is close to the 
\emph{previous} output transition, and the \emph{Charlie effect},
(named after Charles Molnar, who identified its causes in the 70th of the last
century), an increase of the delay happening when the two inputs are
switching in close proximity in the same direction. Clearly, to analyze 
the behavior of the ring, the timing relation of successive
transitions need to be traced throughout the whole
circuit. Since this is outside the scope of static timing analysis techniques,
however, one has to resort to analog simulations e.g.\ via \spice\ \cite{nagel1973spice} 
here. Unfortunately, analog simulation times excessive, even for moderately large 
circuits. This is caused by fact that the dimension of the system of differential 
equations that need to be solved numerically increases with the number of transistors
in a circuit.

\emph{Digital} dynamic timing analysis techniques have been proposed
as a less accurate but much faster alternative for correctness validation
and accurate performance \& power estimation \cite{najm1994survey} of such
circuits, which can even be applied at early design stages. This techniques rests on 
fast and efficiently computable gate delay models, which provide gate delay estimations 
on a per-transition basis.
\emph{Single-history delay models} 
like~\cite{BJV06,FNNS19:TCAD}, which generalize the
popular pure (= constant input-to-output delay) and inertial delay (= constant
delay + too short pulses being removed) models~\cite{Ung71},
have shown their potential for improving accuracy in dynamic timing analysis.

Particularly relevant for us is the
\emph{involution delay model} (IDM) proposed in \cite{FNNS19:TCAD}, which consists
of zero-time boolean gates interconnected by single-input
single-output involution delay channels. By denoting the temporal 
distance of the current input transition to the previous output transition, 
i.e., the previous-output-to-input delay, as $T$, IDM channels are characterized by a
delay function $\delta(T)$ that is a negative involution, in the sense that
$-\delta(-\delta(T))=T$. Note that the dependence on $T$ captures the 
drafting effect introduced in \cite{CharlieEffect}.

Unlike all other existing delay models known so far, the IDM
\emph{faithfully} models glitch propagation in the short-pulse filtration
problem \cite{FNS16:ToC}: it has been proved to be consistent
with physics, in
the sense that a modeled circuit does not exhibit discontinuities at
its output signals when a short pulse at some of its inputs is
shrunk to 0 and hence vanishes. 
Moreover, it has been shown in \cite{FMNNS18:DATE} that one can add 
``delay noise'' to the (obviously deterministic) delay function $\delta(T)$ without sacrificing
faithfulness. As demonstrated in some follow-up work \cite{OS23:arxiv}, delay 
noise can even be used for incorporating substantial PVT variations and even
aging.

The IDM is accompanied by a publicly available timing analysis framework
(the \emph{Involution Tool}~\cite{OMFS20:INTEGRATION}), which allows to
compare the accuracy of different delay models. In particular, it
allows to randomly generate input traces for a given
  circuit, and to evaluate the accuracy of IDM predictions compared to
  \spice-generated transition times and to other digital delay models. 
An experimental evaluation of the modeling accuracy of the IDM 
in \cite{NFNS15:GLSVLSI,OMFS20:INTEGRATION}, using both measurements and simulations, has revealed
very good results for inverter chains and clock trees, albeit
less so for circuits containing also multi-input gates. Note that, in the
case of the clock tree, a speedup of a factor of 250 has been
obtained relative to \spice\ in terms of simulation running times.

The authors argued that the degradation of accuracy in the case of multi-input gates
is primarily due to the IDM's inherent lack of properly covering
output delay variations caused by \emph{multiple input switching} (MIS) in close
temporal proximity~\cite{CGB01:DAC}, i.e., Charlie effects: 
Compared to the \emph{single input switching}
(SIS) case, delays increase/decrease with decreasing
transition separation time on different inputs. Clearly, single-input,
single-output delay channels cannot exhibit such a behavior
at all. This fuels the need for developing alternative digital delay 
models that can cope with MIS effects.

\paragraph*{Related work} MIS effects have of course been addressed in the literature before, with approaches ranging from linear~\cite{SRC15:TDAE} or quadratic~\cite{SKJPC09:ISOCC} fitting over higher-dimensional macromodels~\cite{CS96:DAC} and model representations~\cite{SC06:DATE} to recent machine learning methods~\cite{RS21:TCAD}. However, the resulting models are either empirical or statistical and hence unsuitable as a basis for dynamic digital timing analysis; moreover, they cannot be
proved to be faithful \cite{FNS16:ToC}. 

In \cite{EFS98:ASYNC}, Ebergen, Fairbanks and Sutherland studied the performance of micropipelines,
the control logic of which is made up of a chain of RendezVous elements. Their analysis rests
on a delay model for the RendezVous elements, which relies on the Charlie effect exhibited by the 
constituent Muller \C\ gate. Rather than providing a model that explains this MIS effect, however,
they just considered it as given and used it for analyzing the token propagation performance in successive 
RendezVous elements. The same is true for the already mentioned paper \cite{CharlieEffect} by Winstanley, Garivier and 
Greenstreet.

To the best of our knowledge, the only attempt to develop a delay model that captures MIS effects and is 
suitable for dynamic timing analysis has been provided in \cite{FMOS22:DATE}, where the authors proposed
a hybrid model of a 2-input CMOS \NOR\ gate based on replacing transistors by ideal zero-time switches in the simple RC model shown in \cref{fig:nor_CMOS}.
We recall that a hybrid model describes a hybrid automaton \cite{Hen00}
with several states called modes, in each of which
the behavior of the system outputs is guided by some system of
\emph{ordinary differential equations} (ODEs). The automaton may switch
instantaneously from one mode to another, according to some conditions on the
current state of the system and/or the inputs.
Since a \NOR\ gate has two inputs, it can be described as
a hybrid model with 4 modes, one for each possible digital state of the inputs $(A,B)\in \{(0,0), (0,1), (1,0), (1,1)\}$. 
Each mode is governed by a simple system of constant-coefficient first order
ODEs, which are switched (continuously) upon an input state transition. Whereas this leads to an accurate delay 
model, which has recently even been proved to be also faithful \cite{FFNS23:arxiv}, it fails to capture one of the 
MIS effects, namely, for a rising output transition.

\paragraph*{Main contributions}
(1) A detailed analysis of the shortcomings of \cite{FMOS22:DATE} guided us to a refined approach for developing
hybrid delay models for CMOS gates, where transistors are replaced by \emph{time-varying} resistors. More specifically,
rather than replacing transistors by zero-time switches as in \cite{FMOS22:DATE}, which would be inadequate for
modeling MIS effects in circuits like Muller \C\ gate implementations, we replace (some of) them by continuously
rising resp.\ falling resistors according to the Shichman-Hodges transistor model~\cite{ShichmanHodges}. 
We demonstrate the feasibility of this approach by appying it to both a CMOS \NOR\ gate (which automatically also
covers the ``dual'' \NAND\ gate), and to a Muller \C\ gate. Interestingly, it turned 
out that only the switching-on of the serial transistors needs to be continuous here 
(using the time evolution function $R(t)\sim 1/t + R^{\on}$; all other switchings are instantaneous.

(2) Rather than including the state of the continuously time-varying resistors in the resulting ODE systems
(which would blow up the system dimension), we rely on first-order ODEs with \emph{time-varying coefficients} 
instead. We analytically solved the ODEs for all modes, the complexity of which forced us to use some (close) 
approximations in certain cases, and composed the trajectory functions of different modes, 
like $(0,0) \to (0,1) \to (1,1)$, to determine analytic formulas for all MIS gate delays. 

(3) We provide a procedure for parametrizing our delay model for a given implementation.
Given certain MIS delay values, we used crucial
insights gained from the analytic MIS gate delay formulas derived in (2)
for guiding the process of empirically fitting the various model parameters to match these delay values.
\ifthenelse{\boolean{conference}}{To demonstrate the excellent delay predictions of our model, we show the
results for a CMOS \NOR\ gate in the same 15 nm technology as used in \cite{FMOS22:DATE}. We also performed
analogous validation experiments for a 65 nm \NOR\ gate, as well as for a
Muller \C\ gate, which can only be briefly summarized here due to lack of space.\footnote{In some very recent work,
which cannot be referenced here in order not to break anonymity, we have also augmented our hybrid gate 
delay model for the \NOR\ gate by a RC-type interconnect. Its experimental validation revealed a very good accuracy (in the
\% range) for different wire lengths, driving strenghts, load capacitances, etc.}
}
{
To validate our model, we apply it to a CMOS \NOR\ gate both in a 15 nm technology and in a 65 nm technology,
and to a Muller \C\ gate in 15 nm technology.
}
In all cases, it turned out that the delay predictions of our model match the real 
delays very well, in particular, also in those MIS case where \cite{FMOS22:DATE} fails.

(4) We implemented our hybrid delay model in the Involution Tool~\cite{OMFS20:INTEGRATION}, and experimentally compared the average accuracy and the simulation times of our model to other analog/digital simulations for one of the circuits studied in~\cite{OMFS20:INTEGRATION}. Note that the availability of analytic formulas for the trajectory functions and the gate delays eliminates the need for using a numerical solver in dynamic timing analysis, which is important for small simulation times. As expected, our model outperforms all alternative models known to us in terms of accuracy, without an undue performance penalty.

\emph{Paper organization:}
In \cref{sec:SPIC}, we briefly summarize the results of the analog simulations
used for quantifying MIS delays for the 15~nm CMOS \NOR\ gate in \cite{FMOS22:DATE}, which provide our baseline. \cref{sec:AdvancedModel} introduces our hybrid ODE model. In \cref{Sec:Charlie}, we analytically solve the resulting ODEs and derive gate delay expressions from these solutions. \cref{sec:param} provides
our parametrization procedure, which is applied to \ifthenelse{\boolean{conference}}{our 15~nm CMOS \NOR\ gate simulation data}{both our 15~nm and some 65~nm CMOS \NOR\ gate simulation data} for model validation. Using this parametrization, we finally quantify the average modeling accuracy using the Involution Tool in \cref{sec:modelingaccuracy}. 
\ifthenelse{\boolean{conference}}
{
In \cref{sec:Muller-C}, we sketch the hybrid delay model for a Muller \C\ gate and a glimpse of the obtained results. 
}{
In \cref{sec:Muller-C}, we provide the hybrid delay model for a Muller \C\ gate and some of the results of our validation
experiments. 
}
Some conclusions are provided in \cref{sec:conclusions}.

\begin{table*}
\centering
\caption{Integrals $I_1(t)$, $I_2(t)$, $I_3(t)$ and $U(t)$ for every possible mode switch; $\Delta=t_B-t_A$, and $2R=R_{p_A}+R_{p_B}$.}

\ifthenelse{\boolean{conference}}
{
\scalebox{0.85}
}
{
\scalebox{0.85}
}
{
\begin{tabular}{cccccc}
\hline
Mode                            &  & $I_1(t)= \int_{0}^{t} \frac{\dd s}{R_1(s)+R_2(s)}$                       & $I_2(t)= \int_{0}^{t}\frac{\dd s}{R_3(s)}$                    & $I_3(t)=\int_{0}^{t} \frac{\dd s}{R_4(s)}$                     & $U(t)= \frac{\vdd}{C(R_1(t)+R_2(t))}$                                                         \\ \cline{1-1} \cline{3-6} 
$T^{\uparrow}_{-}$              &  & $\int_{0}^{t} (1/ (\beta_1s+ \frac{\alpha_2}{s-\infty}+2R))\dd s$          & $\int_{0}^{t} (1/(\frac{\alpha_3}{s}+R_{n_A}))\dd s$          & $\int_{0}^{t} (1/(\beta_4(s-\infty)+R_{n_B})) \dd s$           & $\frac{\vdd}{2RC(1+ \frac{\beta_1 t}{2R})}$                                                   \\
$T^{\uparrow \uparrow}_{+}$     &  & $\int_{0}^{t} (1/(\beta_1(s+ \Delta)+\beta_2s+2R))\dd s$                 & $\int_{0}^{t} (1/(\frac{\alpha_3}{s+ \Delta}+R_{n_A}))\dd s$   & $\int_{0}^{t} (1/(\frac{\alpha_4}{s}+R_{n_B})) \dd s$          & $\frac{\vdd}{C((\beta_1 + \beta_2)t + 2R + \beta_1 \Delta)}$                                  \\
$T^{\uparrow}_{+}$              &  & $\int_{0}^{t} (1/ (\frac{\alpha_1}{s-\infty}+ \beta_2s+2R))\dd s$         & $\int_{0}^{t} (1/(\beta_3(s- \infty)+R_{n_A}))  \dd s$        & $\int_{0}^{t} (1/(\frac{\alpha_4}{s}+R_{n_B})) \dd s$          & $\frac{\vdd}{2RC(1+ \frac{\beta_2 t}{2R})}$                                                   \\
$T^{\uparrow \uparrow}_{-}$     &  & $\int_{0}^{t} (1/ (\beta_1 s + \beta_2(s+\Delta)+2R))\dd s$              & $\int_{0}^{t} (1/(\frac{\alpha_3}{s}+R_{n_A})) \dd s$         & $\int_{0}^{t} (1/(\frac{\alpha_4}{s + \Delta}+R_{n_B})) \dd s$   & $\frac{\vdd}{C((\beta_1 + \beta_2)t + 2R + \beta_2 \Delta)}$                                  \\
$T^{\downarrow}_{-}$            &  & $\int_{0}^{t} (1/(\frac{\alpha_1}{s}+ \beta_2 (s- \infty)+2R))\dd s$     & $\int_{0}^{t} (1/(\beta_3s+R_{n_A})) \dd s$                   & $\int_{0}^{t} (1/(\frac{\alpha_4}{s - \infty}+R_{n_B})) \dd s$ & $\frac{\vdd t}{C \beta_2 (t^2 + (\frac{2R}{\beta_2}- \infty)t + \frac{\alpha_1}{\beta_2})}$   \\
$T^{\downarrow \downarrow}_{+}$ &  & $\int_{0}^{t}(1/(\frac{\alpha_1}{s+\Delta}+\frac{\alpha_2}{s}+2R))\dd s$ & $\int_{0}^{t} (1/(\beta_3(s+\Delta)+R_{n_A})) \dd s$          & $\int_{0}^{t} (1/(\beta_4 s + R_{n_B})) \dd s$                 & $\frac{\vdd t(t+ \Delta)}{C(2 R t^2 +(\alpha_1 + \alpha_2 + 2 \Delta R)t + \alpha_2 \Delta)}$ \\
$T^{\downarrow}_{+}$            &  & $\int_{0}^{t}(1/(\frac{\alpha_2}{s}+ \beta_1(s-\infty)+2R))\dd s$         & $\int_{0}^{t} (1/(\frac{\alpha_3}{s- \infty}+R_{n_A})) \dd s$ & $\int_{0}^{t} (1/(\beta_4 s+ R_{n_B})) \dd s$                  & $\frac{\vdd t}{C \beta_1 (t^2 + (\frac{2R}{\beta_1}- \infty)t + \frac{\alpha_2}{\beta_1})}$   \\
$T^{\downarrow \downarrow}_{-}$ &  & $\int_{0}^{t}(1/(\frac{\alpha_1}{s}+\frac{\alpha_2}{s+\Delta}+2R))\dd s$ & $\int_{0}^{t} (1/(\beta_3s +R_{n_A})) \dd s$                  & $\int_{0}^{t} (1/(\beta_4(s+ \Delta)+ R_{n_B})) \dd s$          & $\frac{\vdd t(t+ \Delta)}{C(2 R t^2 +(\alpha_1 + \alpha_2 + 2 \Delta R)t + \alpha_1 \Delta)}$ \\ \hline
\end{tabular}}
\label{tab:T2}
\end{table*}

\section{Multiple Input Switching (MIS)}
\label{sec:SPIC}
In this section, we provide a summary of the MIS effects in CMOS \NOR\
gates 
reported in~\cite{FMOS22:DATE}. Let $t_A$, $t_B$, and $t_O$ denote the points
in time when the analog trajectories of input signals $A$, $B$, and the output
signal $O$ cross the \emph{discretization threshold voltage} $\vth=\vdd/2$,
respectively. Varying $t_A$ and $t_B$ allows to study the gate delay ($t_O-t_A$
resp.\ $t_O-t_B$, depending on the particular output state) over the relative
\emph{input separation time} $\Delta=t_B-t_A$.
	
\ifthenelse{\boolean{conference}}
{}
{
\begin{figure}[t!]
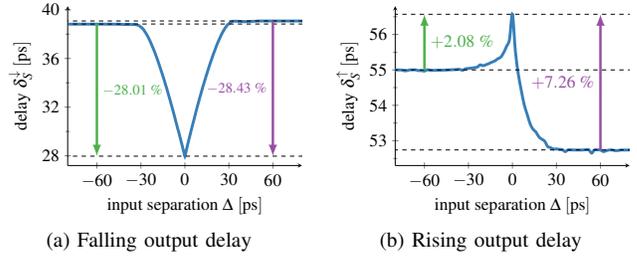

  \centering
  \subfloat[Falling output delay]{
  \ifthenelse{\boolean{conference}}
  {\includegraphics[width=0.35\linewidth]{\figPath{nor2_out_down_charlie_15nm_new.pdf}}}
  {
    \includegraphics[width=0.45\linewidth]{\figPath{nor2_out_down_charlie_15nm_new.pdf}}%
    }
    \label{fig:nor2_out_down_charlie}}
  \hfil
  \subfloat[Rising output delay]{
  
  \ifthenelse{\boolean{conference}}
  {
  \includegraphics[width=0.35\linewidth]{\figPath{nor2_out_up_charlie_15nm.pdf}}
  }
  {
  \includegraphics[width=0.45\linewidth]{\figPath{nor2_out_up_charlie_15nm.pdf}}
  }
    \label{fig:nor2_out_up_charlie}}
  \caption{MIS effect in the CMOS \NOR\ gate, taken from \cite{FMOS22:DATE}.}\label{fig:nor_analog_sim}
\end{figure}
}

In the case of a falling output transition, either the \nmos\ transistor $T_3$
or $T_4$ starts to conduct (is closed), while one of the two \pmos\ transistors
in series stops conducting (is opened). Obviously, closing both $T_3$ and $T_4$
leads to an accelerated discharge of the capacitance $C$ and thus a
(substantial) \textit{speed-up} MIS effect. 
\ifthenelse{\boolean{conference}}
{The dashed red curve in \cref{corFig3}}
{\cref{fig:nor2_out_down_charlie}}
shows the gate delay $\dsd(\Delta)=t_O - \min(t_A, t_B)$, i.e., the
time difference between the threshold crossing of the output and the earlier
input, extracted from analog simulations. The delay for simultaneous transitions
($\Delta=0$) is indeed smaller than on the outskirts.

For rising output transitions, the behavior of the \NOR\ is quite
different. Each falling input transition causes one of the \nmos\ to stop
conducting while simultaneously one of the \pmos\ gets closed. Since there is
only a single path connecting the output to $\vdd$, the shape of the output
signal is essentially independent of $\Delta$; only the position in time varies.
Since the gate only switches after both inputs have changed, the gate delay is
$\dsu(\Delta)=t_O - \max(t_A, t_B)$. The resulting MIS effect is a (moderate)
\textit{slow-down}, i.e., the gate delay increases for $|\Delta| \to 0$, see
\ifthenelse{\boolean{conference}}
{the dashed red curce in \cref{corFig5}.}
{\cref{fig:nor2_out_up_charlie}.}

\section{Our Hybrid Model}
\label{sec:AdvancedModel}

\begin{figure}[t!]
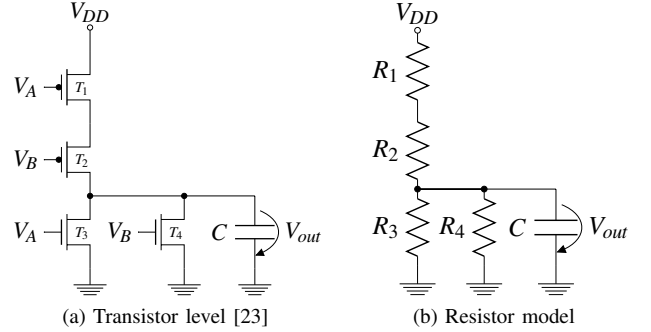

  \centering
  \subfloat[Transistor level~\cite{FMOS22:DATE}]{
  
\ifthenelse{\boolean{conference}}
{
\includegraphics[height=0.30\linewidth]{\figPath{nor_RC.pdf}}
}
{
\includegraphics[height=0.45\linewidth]{\figPath{nor_RC.pdf}}
}  

    \label{fig:nor_CMOS}}
  \hfil
  \subfloat[Resistor model]{
\ifthenelse{\boolean{conference}}
{
 \includegraphics[height=0.30\linewidth]{\figPath{nor_R.pdf}}%
}
{
 \includegraphics[height=0.45\linewidth]{\figPath{nor_R.pdf}}%
}  
    \label{FigureNOR-GATE}}
  \caption{Schematics and resistor model of a CMOS \NOR\ gate.}
\end{figure}

In this section, we will introduce a model (see \cref{FigureNOR-GATE}) that replaces 
transistors by time-varying resistors: The values $R_i(t)$, $i \in \{1,\ldots,4 \}$ thereby
vary between some fixed on-resistance $R_i$ and the off-resistance
$\infty$. This results in a hybrid model with 4 different modes, corresponding to the 4
possible input states $(A,B)\in \{(0,0), (0,1), (1,0), (1,1)\}$.

Crucial for our model is choosing a suitable time evolution of $R_i(t)$ in the
\emph{on-} and \emph{off-mode}, which should facilitate an analytic solution of
the resulting ODE systems \eqref{Eq1} while being reasonably 
close to the physical behavior of a transistor. 
We rely on the very simple Shichman-Hodges 
transistor model~\cite{ShichmanHodges}, which states a quadratic dependence 
of the output current on the input voltage. Approximating the latter by 
$d \sqrt{t-t_0}$ in the operation range close to the threshold voltage $\vth$, 
with $d$ and $t_0$ some fitting parameters, leads to the \emph{continuous resistance model}
\begin{align}
R_j^{\on}(t) &= \frac{\alpha_j}{t-t^{\on}}+R_j; \ t \geq t^{\on}, \label{on_mode}\\
R_j^{\off}(t) &= \beta_j (t-t^{\off}) +R_j; \ t \geq t^{\off}, \label{off_mode}
\end{align}
for some constant slope parameters $\alpha_j$ [\si{\ohm\s}], $\beta_j$
[\si[per-mode=symbol]{\ohm\per\s}], and on-resistance $R_j$ [\si{\ohm}]; $t^{\on}$ resp.\
$t^{\off}$ represent the time when the respective transistor is switched on
resp.\ off.

\ifthenelse{\boolean{conference}}{
\subsection{Model development}
\label{sec:modeldevelopment}
The arguably most natural idea for developing such a model for the \NOR\ gate would 
be to maintain $R_1(t),\dots, R_4(t)$ explicitly in the state of every ODE
system, and switch between those continuously upon a mode switch.
However, to begin with, such a ``full-state model'' would comprise ODE
systems with a 5-dimensional state (output voltage $\vout$ and the 4 resistors)
and thus render finding an analytic solution hopeless.

We therefore use a different approach, namely, incorporating
these resistors only in the \emph{coefficients} of a simple first-order ODE, which
hence become non-constant. Moreover, we employ continuously changing resistors 
(according to \cref{on_mode}) only for switching-on the pMOS transistors. Everything
else will be identical to the ideal switch model (which is obtained by setting $\alpha_j=0$
and $\beta_j=\infty$) already employed in \cite{FMOS22:DATE}. 
}
{
\subsection{Model development}
\label{sec:modeldevelopment}
The arguably most natural idea for developing such a model for the \NOR\ gate would 
be to maintain $R_1(t),\dots, R_4(t)$ explicitly in the state of every ODE
system, and switch between those continuously upon a mode switch.
However, to begin with, such a ``full-state model'' would comprise ODE
systems with a 5-dimensional state (output voltage $\vout$ and the 4 resistors)
and thus render finding an analytic solution hopeless.

We therefore use a different approach, namely, incorporating
these resistors only in the \emph{coefficients} of a simple first-order ODE, which
hence become non-constant. Moreover, we employ continuously changing resistors 
(according to \cref{on_mode}) only for switching-on the pMOS transistors. Everything
else will be identical to the ideal switch model (which is obtained by setting $\alpha_j=0$
and $\beta_j=\infty$) already employed in \cite{FMOS22:DATE}. 
We will argue in \cref{sec:otherapproaches} what guided us to arrive at this model.
}

Applying Kirchhoff's rules to \cref{FigureNOR-GATE} leads to
$ C\frac{\dd \vout}{\dd t} = \frac{\vdd-\vout}{R_1(t)+R_2(t)} -
\frac{\vout}{R_3(t)\ ||\ R_4(t)}\ .$
This can be transformed to the
non-homogeneous ordinary differential equation (ODE) with non-constant coefficients
\begin{equation}
\label{Eq0}
\frac{\dd V_{out}}{\dd t}+\frac{V_{out}}{C\,R_g(t)}=U(t),
\end{equation}
using
$\frac{1}{R_g(t)}=\frac{1}{R_1(t)+R_2(t)}+\frac{1}{R_3(t)}+\frac{1}{R_4(t)}$ and
$U(t)=\frac{V_{DD}}{C(R_1(t)+R_2(t))}$. Note that the entire voltage divider in
\cref{FigureNOR-GATE} is equivalent to an ideal voltage source
$U_0=\vdd\frac{R_3(t)||R_4(t)}{R_1(t)+R_2(t)+R_3(t)||R_4(t)}$ and a serial
resistor $R_g(t)$ sourcing $C$. Consequently, $CU(t)=U_0/R_g(t)$ in \cref{Eq0}
is the short-circuit current, and $CU(t)-\vout/R_g(t)$ the current actually
sourced into $C$. It is well-known that the general
solution of \eqref{Eq0} is
\begin{equation}
\label{Eq1}
    V_{out}(t)= V_0\ e^{-G(t)} + \int_{0}^{t} U(s)\ e^{G(s)-G(t)}\dd s,
\end{equation}
where $V_0=V_{out}(0)$ denotes the initial condition and $G(t) = \int_{0}^{t}
(C\,R_g(s))^{-1} \dd s$.


For simplicity, we will subsequently use the notation $R_1=R_{p_A}$,
$R_2=R_{p_B}$ with the abbreviation $2R=R_{p_A}+R_{p_B}$ for the two pMOS
transistors $T_1$ and $T_2$, and $R_3=R_{n_A}$, $R_4=R_{n_B}$ for the two nMOS
transistors $T_3$ and $T_4$.

\begin{table}[t]
\caption{State transitions and modes. $\uparrow$ and $\uparrow \uparrow$ (resp.\ $\downarrow$ and $\downarrow \downarrow$) represent the first and the second rising (resp.\ falling) input transitions. $+$ and $-$ specify the sign of the switching time $\Delta=t_B-t_A$.}
\scalebox{0.65}
{
\begin{tabular}{ccccccccccc}
\hline
Mode                            &  & Transition                &  & $t_A$       & $t_B$       &  & $R_1$               & $R_2$                & $R_3$                & $R_4$                \\ \cline{1-1} \cline{3-3} \cline{5-6} \cline{8-11} 
$T^{\uparrow}_{-}$              &  & $(0,0) \rightarrow (1,0)$ &  & $0$         & $-\infty$   &  & $on \rightarrow off$ & $on$                 & $off \rightarrow on$ & $off$                \\
$T^{\uparrow \uparrow}_{+}$     &  & $(1,0) \rightarrow (1,1)$ &  & $-|\Delta|$ & $0$         &  & $off$                & $on \rightarrow off$ & $on$                 & $off \rightarrow on$ \\
$T^{\uparrow}_{+}$              &  & $(0,0) \rightarrow (0,1)$ &  & $-\infty$   & $0$         &  & $on$                 & $on \rightarrow off$ & $off$                & $off \rightarrow on$ \\
$T^{\uparrow \uparrow}_{-}$     &  & $(0,1) \rightarrow (1,1)$ &  & $0$         & $-|\Delta|$ &  & $on \rightarrow off$ & $off$                & $off \rightarrow on$ & $on$                 \\
$T^{\downarrow}_{-}$            &  & $(1,1) \rightarrow (0,1)$ &  & $0$         & $-\infty$   &  & $off \rightarrow on$ & $off$                & $on \rightarrow off$ & $on$                 \\
$T^{\downarrow \downarrow}_{+}$ &  & $(0,1) \rightarrow (0,0)$ &  & $-|\Delta|$ & $0$         &  & $on$                 & $off \rightarrow on$ & $off$                & $on \rightarrow off$ \\
$T^{\downarrow}_{+}$            &  & $(1,1) \rightarrow (1,0)$ &  & $-\infty$   & $0$         &  & $off$                & $off \rightarrow on$ & $on$                 & $on \rightarrow off$ \\
$T^{\downarrow \downarrow}_{-}$ &  & $(1,0) \rightarrow (0,0)$ &  & $0$         & $-|\Delta|$ &  & $off \rightarrow on$ & $on$                 & $on \rightarrow off$ & $off$                \\ \hline
\end{tabular}}
\label{tab:T1}
\end{table}

\cref{tab:T1} shows all possible state transitions and the corresponding
resistor time evolution mode switches. Double arrows in the mode switch names
indicate MIS-relevant modes, whereas $+$ and $-$ indicate whether $A$ switched
before $B$ or the other way around.  For instance, assume the system is in state
$(0,0)$ for quite some time in the past, i.e., $A$ and $B$ switched to 0 at time
$t_A= t_B= -\infty$. This causes $R_1$ and $R_2$ to be in the \emph{on-mode},
whereas $R_3$ and $R_4$ are in the \emph{off-mode}. Now assume that, at time
$t_A=0$, $A$ is switched to 1. This switches $R_1$ resp.\ $R_3$ to the
\emph{off-mode} resp.\ \emph{on-mode} at time
$t^{\off}_1 = t^{\on}_3 = t_A = 0$. The corresponding mode switch is
$T_{-}^{\uparrow}$ and reaches state $(1,0)$. Now assume that $B$ is also
switched to 1, at some time $t_B=\Delta>0$.
This causes $R_2$ resp.\ $R_4$ to switch to \emph{off-mode} resp.\
\emph{on-mode} at time $t^{\off}_2= t^{\on}_4 = t_B=\Delta$. The corresponding
mode switch is $T_{+}^{\uparrow\uparrow}$ and reaches state $(1,1)$; note
carefully that the delay is $\Delta$-dependent and hence MIS-relevant.

In \cref{Sec:Charlie}, we will determine analytic formulas for the output voltage trajectories
$V_{out}^{MS}(t)$ given by \cref{Eq1} for every mode switch $MS$ listed in
\cref{tab:T1}. Note that, due to the particular modes of the resistors in each transition, different expressions for $G(t)$
and $U(t)$ according to \cref{tab:T2} will be obtained for every mode switch.

\ifthenelse{\boolean{conference}}
{
We conclude this section by stressing again that the general approach
of replacing transistors with varying resistors advocated in this paper can be readily
applied, and sometimes just transfered, to other gates as well. This is particularly true for the CMOS
\NAND\ gate, which is obtained from the \NOR\ gate in \cref{FigureNOR-GATE}
by replacing nMOS transistors with pMOS and vice versa, and swapping $\vdd$ and $\gnd$. 
Moreover, the hybrid delay model for the Muller \C\ gate in \cref{sec:Muller-C}
confirms that our modeling approach can also be applied easily to other gates
that involve serial or parallel transistors.
}
{
\subsection{Relation to other models}
\label{sec:otherapproaches}
We mentioned already that a 5-dimensional ``full-state model'' would not
allow computing analytic formulas for the trajectories. This is not only
disadvantageous for accuracy validation experiments like the ones presented
in \cref{sec:modelingaccuracy}, as they must resort to a numerical ODE server,
but it also provides no guidance whatsoever for the complex process of model parametrization.
Indeed, in \cref{sec:param}, the availability of analytic expressions of the (inverse) trajectories proved instrumental in developing a successful model parametrization process.

Since we conjectured initially that a full-state model should surpass our
model proposed in \cref{sec:modeldevelopment} 
in terms of modeling accuracy, however, we nevertheless tried hard to 
find a suitable parametrization of the FM for the CMOS \NOR\ gate delays given
in \cref{fig:nor2_out_down_charlie} and \cref{fig:nor2_out_up_charlie}. Whereas
we succeeded to faithfully model the former, we failed to find a 
parametrization that also covers the latter. Given the large parameter space,
we initially blamed our unguided search for the parameters for this failure.

However, as a consequence of another unsuccessful modeling attempt, we eventually 
diagnosed a deeper reason for this failure. In more detail, as a precursor of our 
model, we considered
a version of the model in \cref{sec:modeldevelopment} where both the switching-on
and switching-off resistance of \emph{all} transistors was continuous, according to \cref{on_mode}
and \cref{off_mode}. However, besides considerably complicating the analysis, it eventually 
turned out that this model suffers from the same problem as the full-state model: 
We could not find a 
parametrization that faithfully covers both \cref{fig:nor2_out_down_charlie} and \cref{fig:nor2_out_up_charlie}.

Thanks to the analytic formulas developed for this precursor model, however, we 
eventually realized that the parametrization failures are actually a consequence 
of using the simple resistor model \cref{on_mode} and \cref{off_mode}
in operation regions of a transistor where they do not apply. More specifically,
consider a rising output transition, i.e., switching-on the serial pMOS transistors 
and simultaneously switching-off the parallel nMOS transistors in \cref{FigureNOR-GATE}. 
Since \cref{on_mode} is a convex function, i.e., eventually decreases sub-linearly, 
whereas \cref{off_mode} increases linearly, too much of the current provided by 
the (slowly opening) pMOS transistors is consumed by the (still conducting) nMOS transistors. 
Consequently, $C$ cannot be charged as fast as it should be, no matter how large
$\beta_3, \beta_4 < \infty$ is chosen. Only the immediate switch of the nMOS transistors, 
i.e., choosing $\beta_3=\beta=\infty$, allows to circumvent this problem in our setting.\footnote{We conjecture that the problem
would also disappear if we replaced \cref{on_mode} by a steeper decaying function. However, 
we do not currently know of a suitable candidate that would render solving our non-constant
coefficient ODE \cref{Eq1} possible.} 

We note that the issue explained above also arises in the case of falling output transitions, 
albeit less pronounced: The
current supplied by the (still conducting but serial) switched-off pMOS transistors 
can be reasonably swallowed by the the (slowly opening but parallel) nMOS transistors.
Nevertheless, we witnessed a trace of this problem, by achieving a lower 
parametrization and modeling accuracy of the precursor model also for this case.

One other distinguishing feature of our model is its simplicity, besides the need
to solve a non-constant coefficient ODE. In particular,
whereas the ideal switch model of \cite{FMOS22:DATE} also incorporates 
the voltage at the node between the two pMOS transistors (due to the
parasitic capacitance $C_p$), our model has only one state variable, namely,
$\vout$ (due to the load capacitance $C$). We initially suspected that 
ignoring $C_{int}$ in our model could impair its ability to capture some 
MIS effects, in particular, in the cases of
 falling input transitions $T_{+}^{\downarrow\downarrow}$ and
 $T_{-}^{\downarrow\downarrow}$ (recall \cref{sec:SPIC}) where
the charge stored in $C_{int}$ does affect the gate delay. 
Our results in \cref{sec:param} reveal, however, that this is not the 
case.

In addition, we tried to keep the number of transistors that actually employ a 
continuously varying resistance in our model as small as possible: Only the two pMOS 
transistors $T_1$ and $T_2$ in \cref{FigureNOR-GATE} use \cref{on_mode}, 
all other resistance switches happen instantaneously.

We conclude this section by stressing again that the general approach
of replacing (some) transistors with varying resistors advocated in this paper can be readily
applied, and sometimes just transfered, to other gates as well. This is particularly true for the CMOS
\NAND\ gate, which is obtained from the \NOR\ gate in \cref{FigureNOR-GATE}
by replacing nMOS transistors with pMOS and vice versa, and swapping $\vdd$ and $\gnd$. 
Of course, now the nMOS transistors are the serial ones (where
the continuous resistance models must be used). 
All that is needed to make the results of our analysis matching is to invert the inputs, e.g., by replacing
state $(0,1)$ by $(1,0)$ and to consider $\vdd-\vout(t)$ as the output
trajectory. Moreover, as our model for the Muller \C\ gate in \cref{sec:Muller-C}
confirms, our modeling approach can also be applied easily to other gates, like
\AOI\ (and-or-inverter), that involve serial or parallel transistors.
}

\section{Analytic Solutions}
\label{Sec:Charlie}

In order to verify that the ODE model introduced in \cref{sec:AdvancedModel} faithfully covers all the MIS effects described in \cref{sec:SPIC}, we first derive analytic expressions for the trajectories of $\vout^{MS}(t)$ for every mode switch $MS$ listed in \cref{tab:T1}. Then, we determine the MIS delays for an arbitrary input separation time $\Delta=t_B-t_A$ as follows:
\begin{itemize}
\item Compute $\vout^{T^{\uparrow}_{-}}(\Delta)$, and use it as the initial value for obtaining $\vout^{T^{\uparrow \uparrow}_{+}}(t)$; the sought gate delay is the time until the latter crosses the threshold voltage $\vdd/2$.
\item Compute $\vout^{T^{\downarrow}_{-}}(\Delta)$, and use it as the initial value for obtaining $\vout^{T^{\downarrow \downarrow}_{+}}(t)$; the sought gate delay is the time until the latter crosses the threshold voltage $\vdd/2$.
\end{itemize}
Fortunately, a closer look at \cref{tab:T1} and \cref{tab:T2} shows a symmetry between the pairs of modes $(T^{\uparrow}_{-},T^{\uparrow}_{+})$, $(T^{\uparrow \uparrow}_{+},T^{\uparrow \uparrow}_{-})$, $(T^{\downarrow}_{-},T^{\downarrow}_{+})$, and $(T^{\downarrow \downarrow}_{+},T^{\downarrow \downarrow}_{-})$. Therefore, it is sufficient to derive analytic expressions for the case $\Delta\geq 0$ only.
\ifthenelse{\boolean{conference}}
{}
{
, i.e., to consider the pairs of modes $(T^{\uparrow}_{-}, T^{\uparrow \uparrow}_{+})$ resp.\  $(T^{\downarrow}_{-}, T^{\downarrow \downarrow}_{+})$ listed above.
}
The corresponding formulas for $\Delta < 0$ can be obtained from those by exchanging  $\alpha_1$ and $\alpha_2$ as well as $R_{n_A}$ and $R_{n_B}$, respectively.

\subsection{Rising input transitions}
In order to compute $V_{out}^{T^{\uparrow}_{-}}(t)$, consider the corresponding integrals $I_1(t)$, $I_2(t)$, and $I_3(t)$, as well as $U(t)$ in \cref{tab:T2}. Since $\beta_{1}=\beta_2 = \infty$ and $\alpha_3=\alpha_4=0$, we obtain
\begin{equation}
I_1(t)=I_3(t)=U(t)=0,\qquad
I_2(t)= \frac{t}{R_{n_A}}.\nonumber
\end{equation}
Since $G(t)=(I_1(t)+I_2(t)+I_3(t))/C$, we get $e^{\pm G(t)} = e^{\frac{\pm t}{CR_{n_A}}}$ and $\int_{0}^{t} e^{G(t)} U(t) =0$. With $V_0^{\uparrow} = V_{out}^{T^{\uparrow}_{-}}(0)$ as our initial value, \cref{Eq1} 
finally provides
\begin{align}
V_{out}^{T^{\uparrow}_{-}}(t) = V_{out}^{T^{\uparrow}_{-}}(0) e^{\frac{-t}{C R_{n{A}}}}.
\label{outsig1}
\end{align}

Similarly, for the mode $T^{\uparrow \uparrow}_{+}$, we obtain
\begin{equation}
I_1(t)=U(t)=0,\qquad I_2(t)=\frac{t}{R_{n_A}}, \qquad I_3(t)=\frac{t}{R_{n_B}},\nonumber
\end{equation}
such that $e^{\pm G(t)}= e^{\pm (\frac{1}{CR_{n_A}}+\frac{1}{CR_{n_B}})t}$ and $\int_{0}^{t} e^{G(s)} U(s) ds=0$. Consequently, we obtain
\begin{flalign}
&V_{out}^{T^{\uparrow \uparrow}_{+}}(t) =V_{out}^{T^{\uparrow}_{-}} (\Delta)  e^{- (\frac{1}{CR_{n_A}}+\frac{1}{CR_{n_B}})t},
\label{outsig2}
\end{flalign}
where $V_{out}^{T^{\uparrow}_{-}} (\Delta)$ can be computed via \cref{outsig1}. 

Due to the symmetry mentioned before, we can immediately conclude the following result for 
negative $\Delta$: 
\begin{flalign}
&V_{out}^{T^{\uparrow \uparrow}_{-}}(t) =V_{out}^{T^{\uparrow}_{+}} (\Delta)  e^{- (\frac{1}{CR_{n_A}}+\frac{1}{CR_{n_B}})t},
\label{outsig2_neg}
\end{flalign}
where $V_{out}^{T^{\uparrow}_{+}} (\Delta) = V_{out}^{T^{\uparrow}_{+}}(0) e^{\frac{-|\Delta|}{C R_{n{B}}}}$.

\subsection{Falling input transitions}
In this case, we first need to compute $V_{out}^{T^{\downarrow}_{-}}(t)$. 
Again plugging $\beta_{1}=\beta_2 = \infty$ and $\alpha_3=\alpha_4=0$ in the corresponding
expressions in \cref{tab:T2} provides
\ifthenelse{\boolean{conference}}
{
$I_1(t)=I_2(t)=U(t)=0$ and $I_3(t) = \frac{t}{R_{n_B}}$.
}
{
\begin{equation}
I_1(t)=I_2(t)=U(t)=0,\qquad
I_3(t) = \frac{t}{R_{n_B}}.\nonumber
\end{equation}
}
With $V_{0}^{\downarrow}= V_{out}^{T_{-}^{\downarrow}}(0)$ as our initial condition, 
\cref{Eq1} yields 
\begin{equation}
V_{out}^{T^{\downarrow}_{-}}(t) = V_{out}^{T_{-}^{\downarrow}}(0) e^{\frac{-t}{CR_{n_B}}}.
\label{eq:FirstFall}
\end{equation}

Now, turning our attention to $V_{out}^{T^{\downarrow \downarrow}_{+}}(t)$ confronts us with a more intricate case: Whereas $I_2(t)=I_3(t)=0$ again, evaluating $I_1(t)$ requires us to study the function
\ifthenelse{\boolean{conference}}
{
$f(s)= \frac{1}{\frac{\alpha_1}{s+ \Delta}+ \frac{\alpha_2}{s}+2R}$
}
{
\begin{align}
f(s) &= \frac{1}{\frac{\alpha_1}{s+ \Delta}+ \frac{\alpha_2}{s}+2R},
\label{func:f}
\end{align}
}
as 
\begin{align}
\label{crucial_I1}
&I_1(t)=\int_{0}^{t} f(s) ds, \qquad G(t)=I_1(t)/C,\\
\label{crucial_expint}
&\int_{0}^{t} e^{G(s)} U(s) ds= \frac{\vdd}{C} \int_{0}^{t} e^{\frac{I_1(s)}{C}} f(s) ds.
\end{align}

Computing the above integrals is complicated by the fact that $f(s)$ also involves the parameter
$\Delta$, which prohibits uniform closed-form solutions of \cref{crucial_I1} and \cref{crucial_expint}. We hence need accurate approximations for $f(s)$, which will be different for different ranges of $s$ and $\Delta$, which will in turn depend on the unknown parameters $\alpha_1$, $\alpha_2$ and $R$.
\ifthenelse{\boolean{conference}}
{
Fortunately, a closer look at function $f(t)$ reveals that piecewise approximations can be built on the basis of (i) distinguishing different ranges (cases) for $\Delta$ w.r.t.\ the parameters $\alpha_1$, $\alpha_2$ and $R$, and (ii) splitting the integration interval $[0,t]$ into certain subintervals, which may 
depend on the particular case. Note that this splitting also involves an additional free parameter
$\epsilon=\eta \Delta$ (for some $\eta \in \mathbb{R}$), which takes care of values of $s$ close to $\Delta$.
More specifically, we came up with the following four cases and the 
corresponding approximations for $f(s)$ in the appropriate subintervals:
 
\begin{itemize}
\item \textbf{Case 1:} ($0 \leq \Delta < \frac{\alpha_2}{2R}$)
\begin{align}
f(s) \approx \begin {cases}
\frac{s}{\alpha_2} &   \ \ 0 \leq s < \Delta - \epsilon\\ 
\frac{2s}{\alpha_1 +2 \alpha_2} &  \ \ \Delta- \epsilon \leq s < \Delta + \epsilon \\
\frac{s}{\alpha_1 + \alpha_2} &  \ \ \Delta +\epsilon \leq s < \frac{\alpha_1+ \alpha_2}{2R}\\
\frac{1}{2R} &  \ \ \frac{\alpha_1+ \alpha_2}{2R} \leq s \leq t \nonumber
\end {cases}
\end{align}
\item \textbf{Case 2:} ($\frac{\alpha_2}{2R} \leq \Delta < \frac{\alpha_1+2 \alpha_2}{4R}$)
\begin{align}
f(s) \approx \begin {cases}
\frac{s}{\alpha_2} &   \ \ 0 \leq s < \frac{\alpha_2}{2R} \\ 
\frac{1}{2R} &  \ \ \frac{\alpha_2}{2R} \leq s < \Delta- \epsilon \\
\frac{2s}{\alpha_1 +2 \alpha_2} &  \ \ \Delta -\epsilon \leq s < \Delta +\epsilon\\
\frac{s}{\alpha_1+\alpha_2} &  \ \ \Delta +\epsilon \leq s < \frac{\alpha_1+ \alpha_2}{2R} \\
\frac{1}{2R} &  \ \ \frac{\alpha_1+ \alpha_2}{2R} \leq s \leq t \nonumber
\end {cases}
\end{align}
\item \textbf{Case 3:} ($\frac{\alpha_1+2 \alpha_2}{4R} \leq \Delta < \frac{\alpha_1+ \alpha_2}{2R}$)
\begin{align}
f(s) \approx \begin {cases}
\frac{s}{\alpha_2} &   \ \ 0 \leq s < \frac{\alpha_2}{2R} \\ 
\frac{1}{2R} &  \ \ \frac{\alpha_2}{2R} \leq s < \Delta+ \epsilon \\
\frac{s}{\alpha_1 + \alpha_2} &  \ \ \Delta +\epsilon \leq s < \frac{\alpha_1+ \alpha_2}{2R}\\
\frac{1}{2R} &  \ \ \frac{\alpha_1+ \alpha_2}{2R} \leq s \leq t \nonumber
\end {cases}
\end{align}
\item \textbf{Case 4:} ($\frac{\alpha_1+ \alpha_2}{2R} \leq \Delta < t$)
\begin{align}
f(s) \approx \begin {cases}
\frac{s}{\alpha_2} &   \ \ 0 \leq s < \frac{\alpha_2}{2R} \\ 
\frac{1}{2R} &  \ \ \frac{\alpha_2}{2R} \leq s \leq t \nonumber
\end {cases}
\end{align}
\end{itemize}

To explain how to determine these approximations, we elaborate on how this is done for Case~1; similar arguments can be used to justify the remaining
Cases 2--4. (i) For the range $0 \leq s < \Delta - \epsilon$, which expresses the situation where the integration variable $s$ is fairly smaller than $\Delta$, we
observe $s+\Delta \approx \Delta$, and therefore, $\frac{\alpha_1}{s+ \Delta} \approx \frac{\alpha_1}{\Delta}$. Assuming that the slope parameters $\alpha_1$ and $\alpha_2$ are approximately the same, this leads us to $\frac{\alpha_1}{\Delta}+ \frac{\alpha_2}{s} \approx \frac{\alpha_2}{s}$. Furthermore, since $s < \Delta < \frac{\alpha_2}{2R}$ here, we get $\frac{\alpha_2}{s}+ 2R \approx \frac{\alpha_2}{s}$ and hence $f(s) \approx \frac{s}{\alpha_2}$ as asserted.
(ii) For the range $ \Delta - \epsilon \leq s < \Delta + \epsilon$, where
$s$ is relatively close to $\Delta$, we can substitute $\Delta$ by $s$, which leads to $\frac{\alpha_1}{s+ \Delta} + \frac{\alpha_2}{s} \approx \frac{2 \alpha_1 +\alpha_2}{2s}$. Besides, since $s \approx \Delta < \frac{\alpha_2}{2R} < \frac{\alpha_1 + 2\alpha_2}{4R}$, we get $\frac{2 \alpha_1 + \alpha_2}{2s} + 2R \approx \frac{2 \alpha_1 + \alpha_2}{2s}$, which justifies the asserted
approximation for $f(s)$.
(iii) Turning our attention to the range $\Delta + \epsilon \leq s < \frac{\alpha_1 + \alpha_2}{2R}$, we obtain $\frac{\alpha_1}{s+ \Delta} \approx \frac{\alpha_1}{s}$ and hence $\frac{\alpha_1}{s+ \Delta} + \frac{\alpha_2}{s} \approx \frac{\alpha_1+ \alpha_2}{s}$. Since $s < \frac{\alpha_1 + \alpha_2}{2R}$, this leads to $\frac{\alpha_1+ \alpha_2}{s} + 2R \approx \frac{\alpha_1+ \alpha_2}{s}$ and hence to the asserted approximation.
(iv) Finally, for the remaining range $\frac{\alpha_1 + \alpha_2}{2R} \leq s < t$, the term $2R$ is dominant in the denominator of $f(s)$, hence $f(s) \approx \frac{1}{2R}$.

It is worth pointing out that we had to split the range $0 \leq s < \Delta-\epsilon$ into two parts for Case~2 to improve the approximation accuracy, whereas
we could merge some ranges for Case~4. Whereas we did not bother with determining
analytic bounds for the error of the above first-order Taylor approximations,
given the very small absolut values of $\Delta$ and $s$, we can expect it to be
very small. This is also confirmed by the model validation simulations
in \cref{sec:param}.

The above approximations allow us to easily compute accurate approximations for \cref{crucial_I1} and \cref{crucial_expint}, and thus to determine the output
trajectory $V_{out}^{T^{\downarrow \downarrow}_{+}}(t)$, for every choice of $\Delta$.
More specifically, depending on different Case $k\in\{1,2,3,4\}$, we get $I_1^{(k)}(t) \approx \frac{t}{2R} + i_k$, with
\small
\begin{align}
\label{eq:i1}
&i_1 = \frac{(\Delta- \epsilon)^2}{2\alpha_2} - \frac{(\Delta+ \epsilon)^2}{2(\alpha_1+\alpha_2)} + \frac{4 \epsilon \Delta}{\alpha_1+ 2\alpha_2}- \frac{\alpha_1+ \alpha_2}{8R^2} , \\
\label{eq:i2}
&i_2=\frac{4R(\Delta - \epsilon)-(\alpha_1+ 2 \alpha_2)}{8R^2} - \frac{(\Delta+ \epsilon)^2}{2(\alpha_1+ \alpha_2)} + \frac{4 \epsilon \Delta}{\alpha_1+ 2\alpha_2}, \\
\label{eq:i3}
&i_3=\frac{4R(\Delta + \epsilon)-(\alpha_1+ 2 \alpha_2)}{8R^2}- \frac{(\Delta+ \epsilon)^2}{2(\alpha_1+ \alpha_2)}, \\
\label{eq:i4}
&i_4=-\frac{\alpha_2}{8R^2},
\end{align}
\normalsize
which do not depend on $t$. Furthermore,
\small
\begin{align}
&\int_{0}^{t} e^{G(s)} U(s) ds \approx \frac{\vdd}{C} \cdot \nonumber\\
& \quad\begin {cases}
e^{\frac{i_1}{C}} \bigl (\frac{\int_{0}^{\Delta - \epsilon} s \cdot e^{\frac{s}{2RC}} ds}{\alpha_2}  + \frac{2( \int_{\Delta - \epsilon}^{\Delta+ \epsilon} s \cdot e^{\frac{s}{2RC}} ds)}{\alpha_1+2 \alpha_2}   \\\quad + \frac{\int_{\Delta+ \epsilon}^{\frac{\alpha_1 + \alpha_2}{2R}}s \cdot e^{\frac{s}{2RC}} ds}{\alpha_1+ \alpha_2} + \frac{\int_{\frac{\alpha_1 + \alpha_2}{2R}}^{t} e^{\frac{s}{2RC}} ds}{2R} \bigr ) & \  \text{Case 1} \\ 
e^{\frac{i_2}{C}} \bigl (\frac{\int_{0}^{\frac{\alpha_2}{2R}} s \cdot e^{\frac{s}{2RC}} ds}{\alpha_2}  + \frac{ \int_{\frac{\alpha_2}{2R}}^{\Delta- \epsilon} e^{\frac{s}{2RC}} ds}{2R} +  \frac{2( \int_{\Delta- \epsilon}^{\Delta+ \epsilon} s \cdot e^{\frac{s}{2RC}}ds)}{\alpha_1+ 2 \alpha_2}  \\\quad + \frac{\int_{\Delta+ \epsilon}^{\frac{\alpha_1+ \alpha_2}{2R}} s \cdot e^{\frac{s}{2RC}} ds}{\alpha_1 + \alpha_2} + \frac{\int_{\frac{\alpha_1+ \alpha_2}{2R}}^{t} e^{\frac{s}{2RC}} ds}{2R} \bigr ) &    \ \text{Case 2} \\ 
e^{\frac{i_3}{C}} \bigl (\frac{\int_{0}^{\frac{\alpha_2}{2R}} s \cdot e^{\frac{s}{2RC}} ds}{\alpha_2}  + \frac{ \int_{\frac{\alpha_2}{2R}}^{\Delta+ \epsilon} e^{\frac{s}{2RC}} ds}{2R} +  \frac{ \int_{\Delta+ \epsilon}^{\frac{\alpha_1+ \alpha_2}{2R}} s \cdot e^{\frac{s}{2RC}}ds}{\alpha_1+ \alpha_2} \\\quad + \frac{\int_{\frac{\alpha_1+ \alpha_2}{2R}}^{t} e^{\frac{s}{2RC}} ds}{2R} \bigr ) &   \  \text{Case 3} \\ 
e^{\frac{i_4}{C}} \bigl (\frac{\int_{0}^{\frac{\alpha_2}{2R}} s \cdot e^{\frac{s}{2RC}} ds}{\alpha_2}  + \frac{\int_{\frac{\alpha_2}{2R}}^{t} e^{\frac{s}{2RC}} ds}{2R} \bigr ) &   \ \text{Case 4} \nonumber
\end {cases}
\end{align}
\begin{align}
& = \vdd \cdot e^{\frac{i_k}{C}} \bigl(e^{\frac{t}{2RC}} - \gamma_k  \bigr)  \nonumber
\end{align}
\normalsize
for Case $k \in \{1,2,3,4\}$, where
\small
\begin{flalign} 
\label{eq:gam1}
 \gamma_1 &= \frac{4R^2C}{\alpha_1+ \alpha_2} e^{\frac{\alpha_1+ \alpha_2}{4R^2C}} - \frac{4R^2C}{\alpha_2} &&\nonumber \\
& - \bigl(\frac{2R(\Delta- \epsilon) - 4R^2C}{\alpha_2} - \frac{4R(\Delta - \epsilon) - 8R^2C}{\alpha_1+ 2 \alpha_2} \bigr) e^{\frac{\Delta - \epsilon}{2RC}} &&\nonumber \\
&- \bigl(\frac{4R(\Delta+ \epsilon) - 8R^2C}{\alpha_1 + 2\alpha_2} - \frac{2R(\Delta + \epsilon) - 4R^2C}{\alpha_1+ \alpha_2} \bigr) e^{\frac{\Delta + \epsilon}{2RC}},&&
\end{flalign}
\begin{flalign}
\label{eq:gam2}
\gamma_2 &= \frac{4R^2C}{\alpha_1+ \alpha_2} e^{\frac{\alpha_1+ \alpha_2}{4R^2C}} - \bigl(1 - \frac{4R(\Delta - \epsilon) - 8R^2C}{\alpha_1+ 2 \alpha_2} \bigr) e^{\frac{\Delta - \epsilon}{2RC}} &&\nonumber \\
&- \bigl(\frac{4R(\Delta+ \epsilon) - 8R^2C}{\alpha_1 + 2\alpha_2} - \frac{2R(\Delta + \epsilon) - 4R^2C}{\alpha_1+ \alpha_2} \bigr) e^{\frac{\Delta + \epsilon}{2RC}} &&\nonumber \\
&+ \frac{4R^2C}{\alpha_2} \bigl(e^{\frac{\alpha_2}{4R^2C}} -1 \bigr),&&
\end{flalign}
\begin{flalign}
\label{eq:gam3}
\gamma_3 &= \frac{4R^2C}{\alpha_1+ \alpha_2} e^{\frac{\alpha_1+ \alpha_2}{4R^2C}} - \bigl(1 - \frac{2R(\Delta + \epsilon) - 4R^2C}{\alpha_1+ \alpha_2} \bigr) e^{\frac{\Delta + \epsilon}{2RC}} &&\nonumber \\
&+ \frac{4R^2C}{\alpha_2} \bigl(e^{\frac{\alpha_2}{4R^2C}} -1 \bigr),
\end{flalign}
\begin{flalign}
\label{eq:gam4}
\gamma_4 &= \frac{4R^2C}{\alpha_2} \bigl(e^{\frac{\alpha_2}{4R^2C}}-1 \bigr).&&
\end{flalign}
\normalsize
Combining the above solutions for \cref{crucial_I1} and \cref{crucial_expint}
according to \cref{Eq1}
provides an accurate expression for the output trajectory $V_{out}^{T^{\downarrow \downarrow}_{+}}(t)$:
\begin{equation}
V_{out}^{T^{\downarrow \downarrow}_{+}}(t) \approx 
V_{out}^{T_{-}^{\downarrow}}(\Delta) e^{\frac{-(2i_{k}R+t)}{2RC}}+ \vdd (1- \gamma_k e^{-\frac{t}{2RC}})  \label{eq:bothFall}
\end{equation}
for Case $k\in\{1,2,3,4\}$, where $V_{out}^{T^{\downarrow}_{-}}(\Delta) = V_{out}^{T_{-}^{\downarrow}}(0) e^{\frac{-\Delta}{CR_{n_B}}}$,
$i_1, \ldots, i_4$, $\gamma_1, \ldots, \gamma_4$, are  defined in \cref{eq:FirstFall}, \cref{eq:i1}, ..., \cref{eq:gam4}, respectively. In addition, our symmetry immediately provides us with a trajectory
corresponding to the case of negative $\Delta$:

\begin{equation}
V_{out}^{T^{\downarrow \downarrow}_{-}}(t) \approx 
V_{out}^{T_{+}^{\downarrow}}(\Delta) e^{\frac{-(2i'_{k}R+t)}{2RC}}+ \vdd (1- \gamma'_k e^{-\frac{t}{2RC}})  \nonumber,
\end{equation}
for Case $k\in\{1,2,3,4\}$, where $V_{out}^{T_{+}^{\downarrow}}(\Delta)=V_{out}^{T_{+}^{\downarrow}}(0) e^{-\frac{|\Delta|}{CR_{n_A}}}$ and $i'_1, \ldots, i'_4$, $\gamma'_1, \ldots, \gamma'_4$ are obtained by substituting $\alpha_1$ by $\alpha_2$ and $\Delta$ by $|\Delta|$ in \cref{eq:i1}, ..., \cref{eq:gam4}, respectively.
}
{
Fortunately, a closer look at the function $f(t)$ reveals that piecewise approximations can be built on the basis of (1) distinguishing different ranges (cases) for $\Delta$ w.r.t.\ the parameters $\alpha_1$, $\alpha_2$ and $R$, and (2) splitting the integration interval $[0,t]$ into certain subintervals, which may 
depend on the particular case. Note that this splitting typically also involves an additional free parameter
$\epsilon=\eta \Delta$ (for some $\eta \in \mathbb{R}$), which takes care of values of $s$ close to $\Delta$.
More specifically, we came up with the following four cases and the 
corresponding approximations for $f(s)$ in the appropriate subintervals:
 
\begin{itemize}
\item \textbf{Case 1:} ($0 \leq \Delta < \frac{\alpha_2}{2R}$)
\begin{align}
f(s) \approx \begin {cases}
\frac{s}{\alpha_2} &   \ \ 0 \leq s < \Delta - \epsilon\\ 
\frac{2s}{\alpha_1 +2 \alpha_2} &  \ \ \Delta- \epsilon \leq s < \Delta + \epsilon \\
\frac{s}{\alpha_1 + \alpha_2} &  \ \ \Delta +\epsilon \leq s < \frac{\alpha_1+ \alpha_2}{2R}\\
\frac{1}{2R} &  \ \ \frac{\alpha_1+ \alpha_2}{2R} \leq s \leq t \nonumber
\end {cases}
\end{align}

\item \textbf{Case 2:} ($\frac{\alpha_2}{2R} \leq \Delta < \frac{\alpha_1+2 \alpha_2}{4R}$)
\begin{align}
f(s) \approx \begin {cases}
\frac{s}{\alpha_2} &   \ \ 0 \leq s < \frac{\alpha_2}{2R} \\ 
\frac{1}{2R} &  \ \ \frac{\alpha_2}{2R} \leq s < \Delta- \epsilon \\
\frac{2s}{\alpha_1 +2 \alpha_2} &  \ \ \Delta -\epsilon \leq s < \Delta +\epsilon\\
\frac{s}{\alpha_1+\alpha_2} &  \ \ \Delta +\epsilon \leq s < \frac{\alpha_1+ \alpha_2}{2R} \\
\frac{1}{2R} &  \ \ \frac{\alpha_1+ \alpha_2}{2R} \leq s \leq t \nonumber
\end {cases}
\end{align}

\item \textbf{Case 3:} ($\frac{\alpha_1+2 \alpha_2}{4R} \leq \Delta < \frac{\alpha_1+ \alpha_2}{2R}$)
\begin{align}
f(s) \approx \begin {cases}
\frac{s}{\alpha_2} &   \ \ 0 \leq s < \frac{\alpha_2}{2R} \\ 
\frac{1}{2R} &  \ \ \frac{\alpha_2}{2R} \leq s < \Delta+ \epsilon \\
\frac{s}{\alpha_1 + \alpha_2} &  \ \ \Delta +\epsilon \leq s < \frac{\alpha_1+ \alpha_2}{2R}\\
\frac{1}{2R} &  \ \ \frac{\alpha_1+ \alpha_2}{2R} \leq s \leq t \nonumber
\end {cases}
\end{align}

\item \textbf{Case 4:} ($\frac{\alpha_1+ \alpha_2}{2R} \leq \Delta < t$)
\begin{align}
f(s) \approx \begin {cases}
\frac{s}{\alpha_2} &   \ \ 0 \leq s < \frac{\alpha_2}{2R} \\ 
\frac{1}{2R} &  \ \ \frac{\alpha_2}{2R} \leq s \leq t \nonumber
\end {cases}
\end{align}
\end{itemize}

To explain how to determine these approximations, we elaborate on how this is done for Case~1; similar arguments can be used to justify the remaining
Cases 2--4. (i) For the range $0 \leq s < \Delta - \epsilon$, which expresses the situation where the integration variable $s$ is fairly smaller than $\Delta$, we
observe $s+\Delta \approx \Delta$, and therefore, $\frac{\alpha_1}{s+ \Delta} \approx \frac{\alpha_1}{\Delta}$. Assuming that the slope parameters $\alpha_1$ and $\alpha_2$ are approximately the same, this leads us to $\frac{\alpha_1}{\Delta}+ \frac{\alpha_2}{s} \approx \frac{\alpha_2}{s}$. Furthermore, since $s < \Delta < \frac{\alpha_2}{2R}$ here, we get $\frac{\alpha_2}{s}+ 2R \approx \frac{\alpha_2}{s}$ and hence $f(s) \approx \frac{s}{\alpha_2}$ as asserted.
(ii) For the range $ \Delta - \epsilon \leq s < \Delta + \epsilon$, where
$s$ is relatively close to $\Delta$, we can substitute $\Delta$ by $s$, which leads to $\frac{\alpha_1}{s+ \Delta} + \frac{\alpha_2}{s} \approx \frac{2 \alpha_1 +\alpha_2}{2s}$. Besides, since $s \approx \Delta < \frac{\alpha_2}{2R} < \frac{\alpha_1 + 2\alpha_2}{4R}$, we get $\frac{2 \alpha_1 + \alpha_2}{2s} + 2R \approx \frac{2 \alpha_1 + \alpha_2}{2s}$, which justifies the asserted
approximation for $f(s)$.
(iii) Turning our attention to the range $\Delta + \epsilon \leq s < \frac{\alpha_1 + \alpha_2}{2R}$, we obtain $\frac{\alpha_1}{s+ \Delta} \approx \frac{\alpha_1}{s}$ and hence $\frac{\alpha_1}{s+ \Delta} + \frac{\alpha_2}{s} \approx \frac{\alpha_1+ \alpha_2}{s}$. Since $s < \frac{\alpha_1 + \alpha_2}{2R}$, this leads to $\frac{\alpha_1+ \alpha_2}{s} + 2R \approx \frac{\alpha_1+ \alpha_2}{s}$ and hence to the asserted approximation.
(iv) Finally, for the remaining range $\frac{\alpha_1 + \alpha_2}{2R} \leq s < t$, the term $2R$ is dominant in the denominator of $f(s)$, hence $f(s) \approx \frac{1}{2R}$.

It is worth pointing out that we had to split the range $0 \leq s < \Delta-\epsilon$ into two parts for Case~2 to improve the approximation accuracy, whereas
we could merge some ranges for Case~4. Whereas we did not bother to determine
analytic bounds for the error of the above first-order Taylor approximations,
given the very small absolut values of $\Delta$ and $s$, it is clear that it
should be very small. This is also confirmed by the model validation simulations
in \cref{sec:param}.

The above approximations allow us to easily compute accurate approximations for \cref{crucial_I1} and \cref{crucial_expint}, and thus to determine the output
trajectory $V_{out}^{T^{\downarrow \downarrow}_{+}}(t)$, for every choice of $\Delta$.
More specifically, depending on different Case $k\in\{1,2,3,4\}$, we get $I_1^{(k)}(t) \approx \frac{t}{2R} + i_k$, with
{\small
\begin{align}
\label{eq:i1}
&i_1 = \frac{(\Delta- \epsilon)^2}{2\alpha_2} - \frac{(\Delta+ \epsilon)^2}{2(\alpha_1+\alpha_2)} + \frac{4 \epsilon \Delta}{\alpha_1+ 2\alpha_2}- \frac{\alpha_1+ \alpha_2}{8R^2} , \\
\label{eq:i2}
&i_2=\frac{4R(\Delta - \epsilon)-(\alpha_1+ 2 \alpha_2)}{8R^2} - \frac{(\Delta+ \epsilon)^2}{2(\alpha_1+ \alpha_2)} + \frac{4 \epsilon \Delta}{\alpha_1+ 2\alpha_2}, \\
\label{eq:i3}
&i_3=\frac{4R(\Delta + \epsilon)-(\alpha_1+ 2 \alpha_2)}{8R^2}- \frac{(\Delta+ \epsilon)^2}{2(\alpha_1+ \alpha_2)}, \\
\label{eq:i4}
&i_4=-\frac{\alpha_2}{8R^2},
\end{align}
}
which do not depend on $t$. Furthermore,
\begin{align}
&\int_{0}^{t} e^{G(s)} U(s) ds \approx \frac{\vdd}{C} \cdot \nonumber\\
& \quad\begin {cases}
e^{\frac{i_1}{C}} \bigl (\frac{\int_{0}^{\Delta - \epsilon} s \cdot e^{\frac{s}{2RC}} ds}{\alpha_2}  + \frac{2( \int_{\Delta - \epsilon}^{\Delta+ \epsilon} s \cdot e^{\frac{s}{2RC}} ds)}{\alpha_1+2 \alpha_2}   \\\quad + \frac{\int_{\Delta+ \epsilon}^{\frac{\alpha_1 + \alpha_2}{2R}}s \cdot e^{\frac{s}{2RC}} ds}{\alpha_1+ \alpha_2} + \frac{\int_{\frac{\alpha_1 + \alpha_2}{2R}}^{t} e^{\frac{s}{2RC}} ds}{2R} \bigr ) & \  \text{\small Case 1} \\ 
e^{\frac{i_2}{C}} \bigl (\frac{\int_{0}^{\frac{\alpha_2}{2R}} s \cdot e^{\frac{s}{2RC}} ds}{\alpha_2}  + \frac{ \int_{\frac{\alpha_2}{2R}}^{\Delta- \epsilon} e^{\frac{s}{2RC}} ds}{2R} +  \frac{2( \int_{\Delta- \epsilon}^{\Delta+ \epsilon} s \cdot e^{\frac{s}{2RC}}ds)}{\alpha_1+ 2 \alpha_2}  \\\quad + \frac{\int_{\Delta+ \epsilon}^{\frac{\alpha_1+ \alpha_2}{2R}} s \cdot e^{\frac{s}{2RC}} ds}{\alpha_1 + \alpha_2} + \frac{\int_{\frac{\alpha_1+ \alpha_2}{2R}}^{t} e^{\frac{s}{2RC}} ds}{2R} \bigr ) &    \ \text{\small Case 2} \\ 
e^{\frac{i_3}{C}} \bigl (\frac{\int_{0}^{\frac{\alpha_2}{2R}} s \cdot e^{\frac{s}{2RC}} ds}{\alpha_2}  + \frac{ \int_{\frac{\alpha_2}{2R}}^{\Delta+ \epsilon} e^{\frac{s}{2RC}} ds}{2R} +  \frac{ \int_{\Delta+ \epsilon}^{\frac{\alpha_1+ \alpha_2}{2R}} s \cdot e^{\frac{s}{2RC}}ds}{\alpha_1+ \alpha_2} \\\quad + \frac{\int_{\frac{\alpha_1+ \alpha_2}{2R}}^{t} e^{\frac{s}{2RC}} ds}{2R} \bigr ) &   \  \text{\small Case 3} \\ 
e^{\frac{i_4}{C}} \bigl (\frac{\int_{0}^{\frac{\alpha_2}{2R}} s \cdot e^{\frac{s}{2RC}} ds}{\alpha_2}  + \frac{\int_{\frac{\alpha_2}{2R}}^{t} e^{\frac{s}{2RC}} ds}{2R} \bigr ) &   \ \text{\small Case 4} \nonumber
\end {cases}
\end{align}
\begin{align}
& = \vdd \cdot e^{\frac{i_k}{C}} \bigl(e^{\frac{t}{2RC}} - \gamma_k  \bigr)  \nonumber
\end{align}
for Case $k \in \{1,2,3,4\}$, where
{\small
\begin{flalign} 
\label{eq:gam1}
 \gamma_1 &= \frac{4R^2C}{\alpha_1+ \alpha_2} e^{\frac{\alpha_1+ \alpha_2}{4R^2C}} - \frac{4R^2C}{\alpha_2} &&\nonumber \\
& - \bigl(\frac{2R(\Delta- \epsilon) - 4R^2C}{\alpha_2} - \frac{4R(\Delta - \epsilon) - 8R^2C}{\alpha_1+ 2 \alpha_2} \bigr) e^{\frac{\Delta - \epsilon}{2RC}} &&\nonumber \\
&- \bigl(\frac{4R(\Delta+ \epsilon) - 8R^2C}{\alpha_1 + 2\alpha_2} - \frac{2R(\Delta + \epsilon) - 4R^2C}{\alpha_1+ \alpha_2} \bigr) e^{\frac{\Delta + \epsilon}{2RC}},&&
\end{flalign}
\begin{flalign}
\label{eq:gam2}
\gamma_2 &= \frac{4R^2C}{\alpha_1+ \alpha_2} e^{\frac{\alpha_1+ \alpha_2}{4R^2C}} - \bigl(1 - \frac{4R(\Delta - \epsilon) - 8R^2C}{\alpha_1+ 2 \alpha_2} \bigr) e^{\frac{\Delta - \epsilon}{2RC}} &&\nonumber \\
&- \bigl(\frac{4R(\Delta+ \epsilon) - 8R^2C}{\alpha_1 + 2\alpha_2} - \frac{2R(\Delta + \epsilon) - 4R^2C}{\alpha_1+ \alpha_2} \bigr) e^{\frac{\Delta + \epsilon}{2RC}} &&\nonumber \\
&+ \frac{4R^2C}{\alpha_2} \bigl(e^{\frac{\alpha_2}{4R^2C}} -1 \bigr),&&
\end{flalign}
\begin{flalign}
\label{eq:gam3}
\gamma_3 &= \frac{4R^2C}{\alpha_1+ \alpha_2} e^{\frac{\alpha_1+ \alpha_2}{4R^2C}} - \bigl(1 - \frac{2R(\Delta + \epsilon) - 4R^2C}{\alpha_1+ \alpha_2} \bigr) e^{\frac{\Delta + \epsilon}{2RC}} &&\nonumber \\
&+ \frac{4R^2C}{\alpha_2} \bigl(e^{\frac{\alpha_2}{4R^2C}} -1 \bigr),
\end{flalign}
\begin{flalign}
\label{eq:gam4}
\gamma_4 &= \frac{4R^2C}{\alpha_2} \bigl(e^{\frac{\alpha_2}{4R^2C}}-1 \bigr).&&
\end{flalign}
}
Combining the above solutions for \cref{crucial_I1} and \cref{crucial_expint}
according to \cref{Eq1}
provides an accurate expression for the output trajectory $V_{out}^{T^{\downarrow \downarrow}_{+}}(t)$:

\begin{align}
V_{out}^{T^{\downarrow \downarrow}_{+}}(t) \approx 
V_{out}^{T_{-}^{\downarrow}}(\Delta) e^{\frac{-(2i_{k}R+t)}{2RC}}+ \vdd (1- \gamma_k e^{-\frac{t}{2RC}})  \label{eq:bothFall}
\end{align}
for Case $k\in\{1,2,3,4\}$, where $V_{out}^{T^{\downarrow}_{-}}(\Delta) = V_{out}^{T_{-}^{\downarrow}}(0) e^{\frac{-\Delta}{CR_{n_B}}}$,
$i_1, \ldots, i_4$, $\gamma_1, \ldots, \gamma_4$, are  defined in \cref{eq:FirstFall}, \cref{eq:i1}, ..., \cref{eq:gam4}, respectively. In addition, our symmetry immediately provides us with a trajectory
corresponding to the case of negative $\Delta$:

\begin{equation}
V_{out}^{T^{\downarrow \downarrow}_{-}}(t) \approx 
V_{out}^{T_{+}^{\downarrow}}(\Delta) e^{\frac{-(2i'_{k}R+t)}{2RC}}+ \vdd (1- \gamma'_k e^{-\frac{t}{2RC}})  \nonumber,
\end{equation}
for Case $k\in\{1,2,3,4\}$, where $V_{out}^{T_{+}^{\downarrow}}(\Delta)=V_{out}^{T_{+}^{\downarrow}}(0) e^{-\frac{|\Delta|}{CR_{n_A}}}$ and $i'_1, \ldots, i'_4$, $\gamma'_1, \ldots, \gamma'_4$ are obtained by substituting $\alpha_1$ by $\alpha_2$ and $\Delta$ by $|\Delta|$ in \cref{eq:i1}, ..., \cref{eq:gam4}, respectively.
}
\subsection{Single Input Switching}
Whereas the main focus of our model is the proper modeling of MIS effects, it of course also
needs to handle the ``simple'' single input switching case. This is
done exactly as in the IDM \cite{FNNS19:TCAD}, by continuously switching between the
trajectories of the modes $(0,0)$ and $(1,0)$ (for input A) resp.\ $(0,0)$ and $(0,1)$ 
(for input B).
\ifthenelse{\boolean{conference}}
{
Fortunately, by adjusting  $\Delta$ to be the time difference between the current and the previous transition of the \emph{same} input, we could adapt the above trajectory formulas for the MIS cases to also obtain the ones for the SIS cases. For switching from $(1,0)$ and $(0,0)$, 
for example, it suffices to replace the initial value $V_{out}^{T_{-}^{\downarrow}}(\Delta)$ in $V_{out}^{T_{+}^{\downarrow \downarrow}}(t)$ by the initial value $V_{out}^{T^{\uparrow}_{-}}(\Delta)$, which gives the output voltage at the time the previous switch from $(0,0)$ to $(1,0)$.
}
{
Fortunately, we can adapt the above trajectory formulas for the MIS cases to also obtain the ones for the SIS cases. In fact, in the SIS cases, $\Delta$ just represents the time difference between the current and the previous transition of the \emph{same} input.  For switching from $(1,0)$ and $(0,0)$, for example, it suffices to replace the initial value $V_{out}^{T_{-}^{\downarrow}}(\Delta)$ in $V_{out}^{T_{+}^{\downarrow \downarrow}}(t)$ by the initial value $V_{out}^{T^{\uparrow}_{-}}(\Delta)$, which gives the output voltage at the time the previous switch from $(0,0)$ to $(1,0)$.
}

\section{Parameterization and Modeling MIS Effects}
\label{sec:param}
Since the ultimate goal of our hybrid model is to develop a basis for dynamic
digital timing simulations, our main target is the derivation of explicit analytic formulas for the 
input-to-output delay functions $\delta_M^\downarrow(\Delta)$ and $\delta_M^\uparrow(\Delta)$
for both rising and falling output transitions.
Moreover, we have to answer the question how to determine the parameters
$\alpha_1$, $\alpha_2$, $C$, $R$, $R_{n_A}$, $R_{n_B}$, and $\eta$ for a given
technology. Such a parametrization is of course necessary for checking whether and
how well our new model is capable of faithfully reproducing the MIS effects
described in \cref{sec:SPIC}.

To accomplish the first task, by inverting the explicit formulas obtained for the trajectories $V_{out}^{T^{\uparrow \uparrow}_{+}}(t)$ and $V_{out}^{T^{\uparrow \uparrow}_{-}}(t)$
resp.\ for $V_{out}^{T^{\downarrow \downarrow}_{+}}(t)$ and
$V_{out}^{T^{\downarrow \downarrow}_{-}}(t)$ obtained in the previous section,
we obtain the exact resp.\
approximate analytic expressions for $\delta_{M,+}^\downarrow(\Delta)$ (for $\Delta \geq 0$),
$\delta_{M,-}^\downarrow(\Delta)$ (for $\Delta < 0$) and
$\delta_{M,+}^\uparrow(\Delta)$ (for $\Delta \geq 0$), $\delta_{M,-}^\uparrow(\Delta)$
(for $\Delta < 0$) in terms of the model's parameters given in \cref{thm:delayfunctions}. 

\begin{theorem}[MIS Delay functions]\label{thm:delayfunctions}
For any $0 \leq \Delta \leq \infty$, the MIS delay functions for falling and rising output 
transitions of our model are given by:
\small
\begin{align}
\delta_{M,+}^{\downarrow}(\Delta) = \begin {cases}
 \frac{\ln(2)CR_{n_A}R_{n_B} + \Delta R_{n_B}}{R_{n_A}+R_{n_B}} + \Delta &   \ \ 0 \leq \Delta < \ln(2)CR_{n_A} \nonumber \\ 
 \ln(2)CR_{n_A} &   \ \ \Delta \geq \ln(2)CR_{n_A}
\end {cases}
\end{align}
\begin{align}
\delta_{M,-}^{\downarrow}(\Delta) = \begin{cases}
 \frac{\ln(2)CR_{n_A}R_{n_B} + |\Delta| R_{n_A}}{R_{n_A}+R_{n_B}} + |\Delta| &   \ \ |\Delta| < \ln(2)CR_{n_B} \nonumber \\ 
\ln(2)CR_{n_B} &   \ \ |\Delta| \geq \ln(2)CR_{n_B}
\end {cases}
\end{align}
\normalsize
and, for Case $k\in\{1,2,3,4\}$,
\begin{align}
\delta_{M,+}^{\uparrow}(\Delta) &\approx 
2RC \bigl(\ln(\gamma_k) + \ln(2) \bigr),  \nonumber \\ 
\delta_{M,-}^{\uparrow}(\Delta) &\approx 
2RC \bigl(\ln(\gamma'_k) + \ln(2) \bigr). \nonumber  
\end{align}
$\delta_{M,-}^{\downarrow}(\Delta)$ and $\delta_{M,-}^{\uparrow}(\Delta)$ can be easily obtained by our symmetry.
\end{theorem}


\begin{proof}
We sketch how $\delta_{M,+}^{\downarrow}(\Delta)$ is computed; the expression
for $\delta_{M,-}^{\downarrow}(\Delta)$ is obtained analogously. Recall that 
falling output transitions imply rising input transitions and vice versa. Given
the trajectory $V_{out}^{T^{\uparrow \uparrow}_{+}}(t)$ in \cref{outsig2},
we start out from $V_{out}^{T^{\uparrow}_{-}}(0)=\vdd$ and need to compute the time
$\delta_{M,+}^\downarrow(\Delta)$ when either (i) already the preceding trajectory
$V_{out}^{T^{\uparrow}_{-}}(t)$ or else (ii) $V_{out}^{T^{\uparrow \uparrow}_{+}}(t)$ itself
(which is started at time $\Delta$) hits $\vdd/2$. Note that this reflects the fact that already
the first rising input (at time 0) causes the output to switch to 0. Since all these
trajectories only involve a single exponential, they are easy to invert. It turns out that
case (i) occurs for values $\Delta \geq -\ln(0.5)CR_{n_A}$, case (ii) for smaller $\Delta$.

Similarly, for computing $\delta_{M,+}^{\uparrow}(\Delta)$, we need to consider
the trajectory $V_{out}^{T^{\downarrow \downarrow}_{+}}(t)$. Here, we have
to start out from the initial value $V_{out}^{T^{\downarrow}_{-}}(\Delta)=0$ and 
just need to compute the time $\delta_{M,+}^\uparrow(\Delta)$
until $V_{out}^{T^{\downarrow \downarrow}_{+}}(t)$ hits $\vdd/2$. This reflects the
fact that it is only the second falling input (at time $\Delta$) that causes
the output to switch to 1.
\end{proof}

Whereas the specific initial values $V_{out}^{T^{\uparrow}_{-}}(0)=\vdd$ and $V_{out}^{T^{\downarrow}_{-}}(0)=0$ used in \cref{thm:delayfunctions} are sufficient for evaluating the MIS behavior of our model, we need generalized expressions for the dynamic digital timing simulation experiments in \cref{sec:modelingaccuracy}. In \cref{thm:Edelayfunctions}, we therefore provide the delay functions for arbitrary initial values.

\begin{theorem}
[MIS and SIS delay functions for arbitray initial values]
\label{thm:Edelayfunctions}
For any $0 \leq \Delta \leq \infty$, the extended delay functions for falling and rising output 
transitions of our model, when starting from a given initial value $V_{out}(0)$, are
\small
\begin{align}
\delta_{EM,+}^{\downarrow}(\Delta) = \begin {cases}
- \frac{\ell CR_{n_A}R_{n_B} + \Delta R_{n_B}}{R_{n_A}+R_{n_B}} + \Delta &   \ \ 0 \leq \Delta < -\ell CR_{n_A} \nonumber \\ 
-\ell CR_{n_A} &   \ \ \Delta \geq -\ell CR_{n_A}
\end {cases}
\end{align}
\begin{equation}
\delta_{EM,+}^{\uparrow}(\Delta) \approx 
2RC \bigl( \ln(\frac{2 \vdd \gamma_k - 2 V_{out}(0) e^{-\frac{\Delta}{CR_{n_B}}} e^{-\frac{i_k}{C}}}{\vdd}) \bigr), \nonumber 
\end{equation}
\normalsize
where $\ell=\ln \bigl(\vdd /2V_{out}(0) \bigr)$ and Case $k\in\{1,2,3,4\}$. $\delta_{EM,-}^{\downarrow}(\Delta)$ and $\delta_{EM,-}^{\uparrow}(\Delta)$ can be easily obtained by our symmetry.
\end{theorem}

The delay functions given in \cref{thm:delayfunctions} (that is, in \cref{thm:Edelayfunctions}) 
not only facilitate fast dynamic timing analysis, but are also instrumental for understanding which parameters 
affect the which delay value. In fact, the formulas 
could even be used for explicit parametrization of a given circuit when certain delay
values are known.

\subsection{Parameterization for 15~nm}
In this subsection, we will fit our model to the characteristic
MIS delay values $\ddoD_S(-\infty)$,  $\ddoD_S(0)$, $\ddoD_S(\infty)$ according to \ifthenelse{\boolean{conference}}{\cref{corFig3}}{\cref{fig:nor2_out_down_charlie}} and the corresponding values $\dupD_S(-\infty)$,  $\dupD_S(0)$, $\dupD_S(\infty)$ 
\ifthenelse{\boolean{conference}}{in \cref{corFig5}}{in \cref{fig:nor2_out_up_charlie}} of the CMOS \NOR\ gate implementation described in \cite{FMOS22:DATE}.

Interestingly, our first attempt to simultaneously fit all six delay values
for determining all parameters at once turned out to be naive since impossible. 
To understand why this is the case, note that the
on-resistors of the two nMOS transistors $R_{n_A}$ and $R_{n_B}$ should
roughly be the same. Consequently, we obtain
\ifthenelse{\boolean{conference}}
{
$\frac{\delta_{M,+}^{\downarrow}(\infty)}{\delta_{M,+}^{\downarrow}(0)}= \frac{R_{n_A}+ R_{n_B}}{R_{n_B}} \approx 2.$
}
{
\begin{align}
\frac{\delta_{M,+}^{\downarrow}(\infty)}{\delta_{M,+}^{\downarrow}(0)}= \frac{R_{n_A}+ R_{n_B}}{R_{n_B}} \approx 2. \nonumber
\end{align}
}
Unfortunately, however, the desired ratio is  $\frac{\dsd(-\infty)}{\dsd(0)} \approx \frac{\SI{38}{\ps}}{\SI{28}{\ps}}$, which cannot fit these two values with reasonable choices for $R_{n_A}$ and $R_{n_B}$. As in \cite{FMOS22:DATE}, we fixed this problem by adding (that is, subtracting) a suitably chosen pure delay $\dmin=\SI{18}{\ps}$ (foreseen in the original IDM \cite{FNNS19:TCAD}), which just defers the switching to the new state upon an input transition. This results in an effective ratio of $\frac{\SI{20}{\ps}}{\SI{10}{\ps}}=2$, which could finally be matched by least squares fitting. Of course, when using our model in digital timing analysis, $\dmin$ must be added to the computed delay values.

More specifically, starting out from the desired load capacitance $C$,
we first determined $R_{n_A}$ and $R_{n_B}$ by fitting
$\delta_{M,-}^{\downarrow}(-\infty)$ (or $\delta_{M,+}^{\downarrow}(\infty)$) and $\delta_{M,+}^{\downarrow}(0)$ to match $\ddoD_S(-\infty)-\dmin$ (or $\ddoD_S(\infty)-\dmin$) and $\ddoD_S(0)-\dmin$. Once these parameter values had been obtained, we fixed those and determined
the remaining parameters $R$, $\alpha_1$, $\alpha_2$ and $\eta$ by fitting 
$\delta_{M,-}^{\uparrow}(-\infty)$, $\delta_{M,+}^{\uparrow}(0)$ and $\delta_{M,+}^{\uparrow}(\infty)$ to match $\dupD_S(-\infty)-\dmin$,  $\dupD_S(0)-\dmin$ and $\dupD_S(\infty)-\dmin$.  Note that the same $\dmin=\SI{18}{\ps}$ is used for both rising and falling output transitions. The result of our parametrization \ifthenelse{\boolean{conference}}{}{for the circuit in \cref{sec:SPIC}} is shown in \cref{table:params15}.

\begin{table}[ht]
\ifthenelse{\boolean{conference}}
{
\caption{Model parameter values for the 15~nm CMOS \NOR\ gate from \cite{FMOS22:DATE}.}
}
{
\caption{Model parameter values for the 15~nm CMOS \NOR\ gate used for producing \cref{fig:nor2_out_down_charlie} and \cref{fig:nor2_out_up_charlie}.}
}
\label{table:params15}
\scalebox{0.7}{
\begin{tabular}{cccc}
\hline
\multicolumn{4}{|c|}{Parameters found by fitting the falling output transition case} \\
\multicolumn{1}{|c|}{$R_{n_A}=8.360562682200\; k\Omega$} & \multicolumn{1}{c|}{$R_{n_B}=8.255562682200\; k\Omega$} & \multicolumn{2}{c|}{$C=3.6331599443276\; fF$} \\ \hline\hline
\multicolumn{4}{|c|}{Parameters found by fitting the rising output transition case} \\
\multicolumn{1}{|c|}{$R=6.6999626822002\; k\Omega$} & \multicolumn{1}{c|}{$\alpha_1=0.859\cdot 10^{-7}\; \Omega s$} & \multicolumn{1}{c|}{$\alpha_2=0.268\cdot 10^{-7}\; \Omega s$} & \multicolumn{1}{c|}{$\eta=0.01$} \\ \hline
\end{tabular}}
\end{table}

Utilizing the parameters in \cref{table:params15}, we can finally visualize the delay predictions
of our model. \cref{fig:Charlie15nm} shows the very good fit for both rising and falling
output transitions. We therefore conclude that our new hybrid model fully captures all the MIS effects
introduced in \cref{sec:SPIC}, including the case of $\Delta <0$ for rising
output transitions where the model proposed in \cite{FMOS22:DATE} fails.

\ifthenelse{\boolean{conference}}
{
\begin{figure}[t!]
  \centering
  \subfloat[Falling output delay]{
    \includegraphics[width=0.35\linewidth]{\figPath{hm_falling_output_new.pdf}}%
    \label{corFig3}}
  \hfil
  \subfloat[Rising output delay]{
    \includegraphics[width=0.35\linewidth]{\figPath{hm_rising_output_complete.pdf}}%
    \label{corFig5}}
  \caption{Computed ($\delta_M^{\uparrow}(\Delta)$) and measured ($\delta_S^{\uparrow}(\Delta)$) MIS delays for the 15~nm
CMOS \NOR\ gate from \cite{FMOS22:DATE}.}\label{fig:Charlie15nm}
\end{figure}
}
{
\begin{figure}[t!]
  \centering
  \subfloat[Falling output delay]{
    \includegraphics[width=0.45\linewidth]{\figPath{hm_falling_output_new.pdf}}%
    \label{corFig3}}
  \hfil
  \subfloat[Rising output delay]{
    \includegraphics[width=0.45\linewidth]{\figPath{hm_rising_output_complete.pdf}}%
    \label{corFig5}}
  \caption{Computed ($\delta_M^{\uparrow}(\Delta)$) and measured ($\delta_S^{\uparrow}(\Delta)$) MIS delays for the 15~nm
CMOS \NOR\ gate.}\label{fig:Charlie15nm}
\end{figure}
}

\ifthenelse{\boolean{conference}}
{}
{
As a final remark, we note that some of the modeling inaccuracies visible in \cref{corFig5}
are caused by the approximation errors introduced by
our solutions of \cref{crucial_I1} and \cref{crucial_expint}.
One is the lack of perfectly fitting the actual delay with the computed one for $\Delta=0$, which looks quite bad in the figure but actually causes a relative error of
about 0.95~\% only. Other instances are the small dent shapes around $\Delta \approx \pm 7.64$~ps, which are caused by the approximation error at the ending boundary of the range in Case 3. It could easily be circumvented by slightly moving the border between Case 3 and Case 4 to a smaller value, i.e., to $\frac{\alpha_1 + \alpha_2}{2R}-\varepsilon$ for some
$\varepsilon >0$. Since the induced error is marginal, however, we did not bother with further complicating our analysis.
}

\subsection{Other parameterizations}
A crucial feature of any model is wide applicability. Ideally, our hybrid delay model should
be applicable to any CMOS technology, for any supply voltage, temperature, age etc., in the
sense that it is possible to determine a parametrization that allows our model to match the
MIS delays of any given CMOS \NOR\ implementation. Overall, it is reasonable
to conjecture that our model is applicable whenever the Shichman-Hodges transistor 
model~\cite{ShichmanHodges}, i.e., \cref{on_mode} and \cref{off_mode}, reasonably applies.
\ifthenelse{\boolean{conference}}
{

Whereas it is of course impossible for us to prove such a claim, we can demonstrate that this 
is the case for some quite different technology and operation conditions, namely
the UMC \SI{65}{\nm} technology with $\vdd=\SI{1.2}{V}$ supply voltage and a larger load
capacitance $C$.\footnote{We note that we played with several different conditions and
configurations, which all confirmed that the parameters of our model can be easily matched.}
The red dashed curves in \cref{fig:Charlie65nmR} (falling output) resp.\ \cref{fig:Charlie65nmF} 
(rising output), which correspond to \cref{corFig3} resp.\ \cref{corFig5}, show the gate delays 
depending on the input separation time $\Delta$ obtained via \spice\ simulations. Parametrizing our model for these \SI{65}{\nm} MIS delays turned out to be easy by following the approach already used for \SI{15}{\nm}. Note carefully, however, that we had to use a different pure delay of $\dmin=\SI{10.8}{\ps}$ here, since $\frac{\dsd(-\infty)}{\dsd(0)} \approx \frac{\SI{222}{\ps}}{\SI{116}{\ps}}$. \cref{table:params65} provides the resulting list of parameters. The blue curves in \cref{fig:Charlie65nmR} and \cref{fig:Charlie65nmF} illustrate the computed delay of our model: We see an ideal fitting for the case of falling output transition as well as for the case of rising output transitions at marginal $\Delta$ values. Particularly remarkable is the negligible absolute error value of $0.3\%$ for the mismatch associated with $\Delta=0$.

\begin{table}[ht] 
\caption{Model parameter values for the \SI{65}{\nm} CMOS \NOR\ gate.}
\label{table:params65}
\scalebox{0.7}{
\begin{tabular}{cccc}
\hline
\multicolumn{4}{|c|}{Parameters found by fitting the falling output transition case} \\
\multicolumn{1}{|c|}{$R_{n_A}=8.409562682200\; k\Omega$} & \multicolumn{1}{c|}{$R_{n_B}=8.285562682200\; k\Omega$} & \multicolumn{2}{c|}{$C=30.6331599443276\; fF$} \\ \hline\hline
\multicolumn{4}{|c|}{Parameters found by fitting the rising output transition case} \\
\multicolumn{1}{|c|}{$R=5.1916426822002\; k\Omega$} & \multicolumn{1}{c|}{$\alpha_1=0.959\cdot 10^{-7}\; \Omega s$} & \multicolumn{1}{c|}{$\alpha_2=0.273\cdot 10^{-7}\; \Omega s$} & \multicolumn{1}{c|}{$\eta=0.01$} \\ \hline
\end{tabular}}
\end{table}

\begin{figure}[t!]
  \centering
  \subfloat[Falling output delay]{
 \includegraphics[width=0.35\linewidth]{\figPath{nor2_out_down_charlie.pdf}}
    \label{fig:Charlie65nmR}}
  \hfil
  \subfloat[Rising output delay]{
\includegraphics[width=0.35\linewidth]{\figPath{nor2_out_up_charlie.pdf}}%
    \label{fig:Charlie65nmF}}
  \caption{Computed ($\delta_M^{\uparrow}(\Delta)$) and measured ($\delta_S^{\uparrow}(\Delta)$) MIS delays for \SI{65}{\nm} technology.}\label{fig:Charlie65nm}
\end{figure}
}
{
Whereas it is of course impossible for us to prove such a claim, we can demonstrate that this 
is the case for some quite different technology and operation conditions, namely
the UMC \SI{65}{\nm} technology with $\vdd=\SI{1.2}{V}$ supply voltage and a larger load
capacitance $C$.\footnote{We note that we played with several different conditions and
configurations, which all confirmed that the parameters of our model can be easily matched.}
The red dashed curves in \cref{fig:Charlie65nmR} (falling output) resp.\ \cref{fig:Charlie65nmF} 
(rising output), which correspond to \cref{corFig3} resp.\ \cref{corFig5}, show the gate delays 
depending on the input separation time $\Delta$ obtained via \spice\ simulations. 

Parametrizing our model for these \SI{65}{\nm} MIS delays turned out to be remarkable easy:
Exactly as for our \SI{15}{\nm} technology, by initially fixing some value for the load capacitance $C$ and trying to match $\delta_{M,-}^{\downarrow}(-\infty)$ (or $\delta_{M,+}^{\downarrow}(\infty)$) and $\delta_{M,+}^{\downarrow}(0)$ with $\ddoD_S(-\infty)-\dmin$ (or $\ddoD_S(\infty)-\dmin$) and $\ddoD_S(0)-\dmin$, we determined accurate values for $R_{n_A}$ and $R_{n_B}$. After fixing the latter, we obtained the values of the remaining parameters by fitting the remaining delay values.  Note carefully, however, that we had to use a different pure delay $\dmin=\SI{10.8}{\ps}$ here, since $\frac{\dsd(-\infty)}{\dsd(0)} \approx \frac{\SI{222}{\ps}}{\SI{116}{\ps}}$. \cref{table:params65} provides the resulting list of parameters. Not surprisingly, since the gate delay is considerably higher than for our \SI{15}{\nm} data, we indeed face a significantly larger value for the load capacitance $C$ also in our model.

\begin{figure}[t!]
  \centering
  \subfloat[Falling output delay]{
 \includegraphics[width=0.45\linewidth]{\figPath{nor2_out_down_charlie.pdf}}
    \label{fig:Charlie65nmR}}
  \hfil
  \subfloat[Rising output delay]{
\includegraphics[width=0.45\linewidth]{\figPath{nor2_out_up_charlie.pdf}}%
    \label{fig:Charlie65nmF}}
  \caption{Computed ($\delta_M^{\uparrow}(\Delta)$) and measured ($\delta_S^{\uparrow}(\Delta)$) MIS delays for \SI{65}{\nm} technology.}\label{fig:Charlie65nm}
\end{figure}

\begin{table}[ht] 
\caption{Model parameter values for the \SI{65}{\nm} CMOS \NOR\ gate used for producing the red curves in \cref{fig:Charlie65nm}.}
\label{table:params65}
\scalebox{0.7}{
\begin{tabular}{cccc}
\hline
\multicolumn{4}{|c|}{Parameters found by fitting the falling output transition case} \\
\multicolumn{1}{|c|}{$R_{n_A}=8.409562682200\; k\Omega$} & \multicolumn{1}{c|}{$R_{n_B}=8.285562682200\; k\Omega$} & \multicolumn{2}{c|}{$C=30.6331599443276\; fF$} \\ \hline\hline
\multicolumn{4}{|c|}{Parameters found by fitting the rising output transition case} \\
\multicolumn{1}{|c|}{$R=5.1916426822002\; k\Omega$} & \multicolumn{1}{c|}{$\alpha_1=0.959\cdot 10^{-7}\; \Omega s$} & \multicolumn{1}{c|}{$\alpha_2=0.273\cdot 10^{-7}\; \Omega s$} & \multicolumn{1}{c|}{$\eta=0.01$} \\ \hline
\end{tabular}}
\end{table}

Utilizing the parameters in \cref{table:params65}, we finally obtained the blue curves in \cref{fig:Charlie65nmR} and \cref{fig:Charlie65nmF}, which illustrate the computed delays of our model for the \SI{65}{\nm} technology. It is apparent that they match the real delays very well: We see an ideal fitting for the case of falling output transition as well as for the case of rising output transitions at marginal $\Delta$ values. Moreover, considerably promising is the negligible absolute error value of $0.3\%$ for the mismatch associated with $\Delta=0$.
}

\section{Modeling Accuracy Experiments}
\label{sec:modelingaccuracy}
\ifthenelse{\boolean{conference}}
{
In this section, we compare the modeling accuracy of our model to the ideal 
switch hybrid model \cite{FMOS22:DATE}, the IDM and to classic inertial delays,
using the publicly available Involution Tool \cite{OMFS20:INTEGRATION}.
Albeit it performs dynamic digital timing simulation in VHDL, it also
supports delay models implemented in Python. We hence implemented our model,
that is, the delay functions given in \cref{thm:Edelayfunctions}, in Python.
}
{
In this section, we experimentally compare the modeling accuracy of our new model to the ideal 
switch hybrid model \cite{FMOS22:DATE}, the IDM and to classic inertial delays,
using the publicly available Involution Tool \cite{OMFS20:INTEGRATION}.
Albeit it performs dynamic digital timing simulation in VHDL, it also
supports delay models implemented in Python. We hence implemented our model,
that is, the delay functions given in \cref{thm:Edelayfunctions}, in Python.
}

For our experimental evaluation, we used the same setup as in \cite{FMOS22:DATE}, namely 
the \SI{15}{\nm} Nangate Open Cell Library
featuring FreePDK15$^\text{TM}$ FinFET models~\cite{Nangate15}
($\vdd=\SI{0.8}{\V}$) that has also been used in \cref{sec:SPIC}. 
Based on a Verilog description of our \NOR\ gate, 
we performed optimization, placement 
and routing by utilizing the Cadence tools Genus and Innovus (version 19.11).  
We also extracted the parasitic networks from the final 
layout to obtain \spice\ models.
These models allowed us to perform simulations with Spectre (version 19.1),
which are used to produce golden reference digital signal traces, by
recording their $\vdd/2$ crossing times. All simulations in our delay model
used the final parameters from \cref{table:params15}. For 
the ideal switch model and the IDM Exp-channel, the parameters reported in
\cite{FMOS22:DATE} were used.

In order to quantify and compare the typical (average) modeling accuracy 
of our competing delay models, we stimulated our \NOR\ circuit with randomly
generated waveforms. Our experiments targeted both fast (100/50) and slow 
(200/100) pulse trains on every input, with (LOCAL) and without (GLOBAL) 
concurrent transitions. For example, the configuration \emph{100/50 - LOCAL} 
of the Involution Tool generates transitions independently on both input A and B, 
according to a normal distribution with $\mu = \SI{100}{\ps}$ and 
$\sigma = \SI{50}{\ps}$. The configuration \emph{200/100 - GLOBAL}
generates transitions either on input A or on B, with $\mu = \SI{200}{\ps}$ and 
$\sigma = \SI{100}{\ps}$. Each simulation run consisted of 500 transitions and has been 
repeated 200 times.

As our modeling accuracy comparison metric, we used the area under the deviation trace, which is the
absolute value of the difference between the \spice\ trace and the trace 
generated by the respective delay model integrated over time. However,
since the absolute values largely depend on the number of transitions, and are 
therefore meaningless per se, we normalized them with respect to the inertial 
delays, which is our baseline model.
Consequently, lower bars indicate less area under the deviation trace and 
hence better results. \cref{fig:nor_results} shows the results of our experiments. It is apparent that our model
outperforms the ideal switch hybrid model \cite{FMOS22:DATE} in the case of fast pulse trains,
where the increased modelling accuracy of the falling output MIS delay comes into effect: Around
$5 \%$ is gained in the case of \emph{100/50 - LOCAL}.
Whereas this improvement appears to be small, it needs to be stressed that it is
the \emph{average} accuracy that we are evaluating here: Since the randomly generated inputs are unlikely to create many  transitions very close to each other
(i.e., small $\Delta$), where the superiority of our model w.r.t.\ MIS effects for rising output transitions 
would kick in,
one cannot expect much improvement here. In applications like \cite{CharlieEffect}, however, 
pulse trains containing such transitions do occur.

\begin{figure}[t]
	\centering
	\ifthenelse{\boolean{conference}}
{
\includegraphics[width=0.9\linewidth]{\figPath{results_nor_gate_15nm.pdf}}
	\caption{\label{fig:nor_results} Accuracy of inertial delay,
		Exp-Channel, ideal switch model and our new model, compared to \spice\ 
		simulations of a \NOR\ gate.}
}
{
\includegraphics[width=.96\linewidth]{\figPath{results_nor_gate_15nm.pdf}}
	\caption{\label{fig:nor_results} Accuracy of inertial delay,
		Exp-Channel, ideal switch model and our new model, compared to \spice\ 
		simulations of a \NOR\ gate. Lower bars indicate better results.
		}
}
\end{figure}	


We conclude this section with noting that the simulation times (in the few
second-range) for all our models turned out to be similar. More specifically, the inertial
delay model (which is natively implemented in ModelSim) is only around 33\% faster than both 
the IDM model and the two hybrid models in all our configurations. Given that both hybrid models
have been implemented in Python, we can safely say that predicting gate delays using 
our new model does not incur a significant overhead in terms of simulation running times.
Whereas it appears that the simulation times in \spice\ also match
the ones of ModelSim in configurations like \emph{100/50 - LOCAL} and the ones
of our hybrid models in \emph{5000/5 - GLOBAL}, this is only due to the fact that
the simulated circuit is so small. As already mentioned in \cref{sec:intro}, the
\spice\ simulation times quickly go up with the number of transistors in the circuit.

\section{Hybrid Delay Models for Other Gates}
\label{sec:Muller-C}
In this section, we will demonstrate that our general approach can be applied to other gates as well. We mentioned
already at the end of \cref{sec:modeldevelopment} that our results for the \NOR\ gate can be transferred to the
case of a \NAND\ gate by simply swapping \nmos\ transistors with \pmos\ transistors and vice versa, as well as swapping the supply voltage ($\vdd$ and $\gnd$). In addition, we mentioned that, in some recent work that could not be referenced
for anonymity reasons, we extended our model for the \NOR\ gate by adding an RC interconnect and experimentally
verified its  very good accuracy. 

More generally, we believe that it is easy to apply our modeling approach to any CMOS gate
that involves serial and parallel transistors only, like Muller \C\ gates \C and \AOI\ (and-or-inverter) gates. 
We will demonstrate this for the Muller \C\ gate, which plays a prominent role in many asynchronous circuit designs, 
like the one \cite{CharlieEffect} already mentioned in \cref{sec:intro}.

We will apply our approach to the CMOS implementation of \C\ shown in \cref{fig:C_trans_impl}, which involves a state
keeper element made up of a loop of two inverters. Whereas this seems to complicate our modeling at first sight,
it turns out that we can safely drop it from our considerations altogether: The (ideal) load capacitance $C$ 
in the corresponding resistor model \cref{fig:C_Model_impl} acts as the state keeper for $\vout$ in the case
the output is in a high-impedance state, i.e., when both at least one of $P1$ and $P2$ and at least one of
$N1$ and $N2$ are switched off. In order to accommodate the negation of the output, we just let
$R_1$ and $R_2$ correspond to the nMOS transistors $N_2$ and $N_1$, and $R_3$ and $R_4$ to the pMOS transistors 
$P_1$ and $P_2$. 

Similar to the \NOR\ gate, we use continuously varying resistors according to \cref{on_mode} for the switching-on and instantaneous switching-off, but this time for all transistors. Applying Kirchhoff's rules to \cref{fig:C_Model_impl} 
leads to the first-order non-homogeneous ordinary differential equation (ODE) with non-constant coefficients
\begin{equation}
\label{Eq:C_diff}
\frac{\dd V_{out}}{\dd t}+\frac{V_{out}}{C\,R_g(t)}=U(t),
\end{equation}
where $1/R(g(t))=\frac{1}{R_1(t)+R_2(t)} +\frac{1}{R_3(t)+R_4(t)}$ and $U(t)=\frac{\vdd}{C(R_1(t)+R_2(t))}$. 
It is not difficult to check that a similar solution approach as described in \cref{Sec:Charlie} 
leads to the following expressions for the output voltage, for rising resp.\ falling input transitions
and $\Delta \geq 0$, corresponding to \cref{outsig1} and \cref{outsig2} resp.\ \cref{eq:FirstFall} and \cref{eq:bothFall}:
\begin{align}
V_{out}^{T^{\uparrow}_{-}}(t) = V_{out}^{T^{\uparrow}_{-}}(0), \qquad V_{out}^{T^{\uparrow \uparrow}_{+}}(t) =& V_{out}^{T^{\uparrow}_{-}}(\Delta) e^{\frac{-(2 \bar{i_k} R_n +t)}{2R_nC}} \nonumber \\ & + V_{DD}(1- \bar{\gamma_k}e^{\frac{-t}{2R_nC}}), \nonumber \\
V_{out}^{T^{\downarrow}_{-}}(t) = V_{out}^{T^{\downarrow}_{-}}(0), \qquad V_{out}^{T^{\uparrow \uparrow}_{+}}(t) =&V_{out}^{T^{\downarrow}_{-}}(\Delta) e^{\frac{-(2 \bar{i'_k} R_p +t)}{2R_pC}}, \nonumber
\label{C_Voltage}
\end{align}
with $V_{out}^{T^{\uparrow}_{-}}(0)$ and $V_{out}^{T^{\downarrow}_{-}}(0)$ as the initial values of the transitions $(0,0) \rightarrow (1,0)$, and $(1,1) \rightarrow (0,1)$, respectively. Herein, $\bar{i_k}$ and $\bar{\gamma_k}$ are defined exactly as in \cref{eq:i1} to \cref{eq:gam1} with substituting $2R$ by $2R_n=R_{n_{A}}+R_{n_{B}}$. Similarly, $\bar{i'_k}$ and $\bar{\gamma'_k}$ are obtained by replacing $\alpha_1$, $\alpha_2$ with $\alpha_4$, $\alpha_3$ in \cref{eq:i1} to \cref{eq:gam1}. Finally, the following theorem, whose proof is very similar to the one for \cref{thm:Edelayfunctions}, provides the delay formulas for the \C\ gate.
\begin{theorem}
[MIS and SIS delay functions for the \C\ gate, for arbitray initial values]
\label{thm:Edelayfunctions_C}
For any $0 \leq \Delta \leq \infty$, the delay functions for falling and rising output 
transitions of our model, when starting from a given initial value $V_{out}(0)$, are
\small
\begin{align}
&\delta_{EM,+}^{\uparrow \ \C\ }(\Delta) \approx 
2R_nC \bigl( \ln(\frac{2 ( \vdd \bar{\gamma_k} - V_{out}^{T^{\uparrow}_{-}}(\Delta) e^{-\frac{\bar{i_k}}{C}})}{V_{DD}} \bigr), \nonumber  \\
&\delta_{EM,+}^{\downarrow \ \C\ }(\Delta) \approx 
2R_pC \bigl( \ln(\frac{2 V_{out}^{T^{\downarrow}_{-}}(\Delta)}{V_{DD}})  - \frac{\bar{i'_k}}{C} \bigr). \nonumber
\end{align}
\normalsize
\end{theorem}
As for the \NOR\ gate, the trajectory and delay formulas for $\Delta <0$ can be easily obtained by the symmetry mentioned in \cref{Sec:Charlie}.

\begin{figure*}[t]
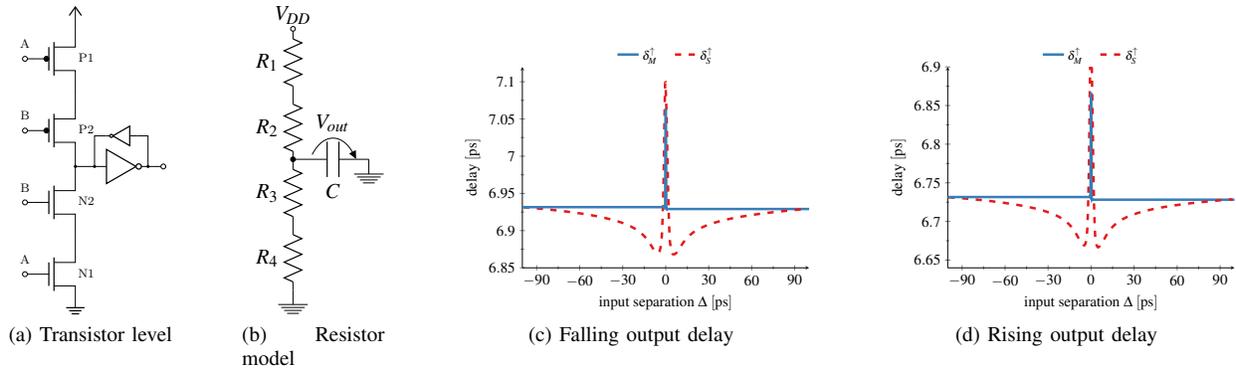

  \centering
\ifthenelse{\boolean{conference}}
{
  \subfloat[Transistor level]{
\includegraphics[height=0.18\linewidth]{\figPath{Martin_C_element.pdf}}
    \label{fig:C_trans_impl}}
  \hfil
  \subfloat[Resistor model]{
 \includegraphics[height=0.18\linewidth]{\figPath{C_Model.pdf}}%
    \label{fig:C_Model_impl}}
      \hfil
  \subfloat[Falling output delay]{
 \includegraphics[height=0.18\linewidth]{\figPath{Cgate_RIS_INPUT.pdf}}%
    \label{fig:charlieC_RIS}} 
      \hfil
  \subfloat[Rising output delay]{
 \includegraphics[height=0.18\linewidth]{\figPath{Cgate_Fall_INPUT.pdf}}%
    \label{fig:charlieC_FALL}}
}
{
  \subfloat[Transistor level]{
\includegraphics[height=0.23\linewidth]{\figPath{Martin_C_element.pdf}}
    \label{fig:C_trans_impl}}
  \hfil
  \subfloat[Resistor model]{
 \includegraphics[height=0.23\linewidth]{\figPath{C_Model.pdf}}%
    \label{fig:C_Model_impl}}
      \hfil
  \subfloat[Falling output delay]{
 \includegraphics[height=0.20\linewidth]{\figPath{Cgate_RIS_INPUT.pdf}}%
    \label{fig:charlieC_RIS}} 
      \hfil
  \subfloat[Rising output delay]{
 \includegraphics[height=0.20\linewidth]{\figPath{Cgate_Fall_INPUT.pdf}}%
    \label{fig:charlieC_FALL}}
}
  \caption{CMOS \SI{15}{\nm} technology \C\ gate implementation, along with the predicted ($\delta_M^{\uparrow}(\Delta)$) and measured ($\delta_S^{\uparrow}(\Delta)$) MIS delays.}
\end{figure*}

We demonstrate the good accuracy of this hybrid delay model by comparing the model predictions and the \spice\ simulation data for a $15$nm CMOS technology \C\ gate. For model parametrization, we could re-use the procedure for the \NOR\ gate, which
provided the parameters listed in in \cref{T:Cgate}. The comparison results are shown in \cref{fig:charlieC_RIS} and \cref{fig:charlieC_FALL}.

\begin{table}[ht] 
\caption{Model parameter values for the \SI{15}{\nm} \C\ gate that produced the dahed red curves in \cref{fig:charlieC_RIS} and \cref{fig:charlieC_FALL}.}
\label{T:Cgate}
\scalebox{0.7}{
\begin{tabular}{cccc}
\hline
\multicolumn{4}{|c|}{Parameters found by fitting the falling output transition case} \\
\multicolumn{1}{|c|}{$R_n=3.002226\; k\Omega$} & \multicolumn{1}{c|}{$\alpha_1=0.161e-8\cdot 10^{-8}\; \Omega s$} & \multicolumn{1}{c|}{$\alpha_2=0.158e-8\cdot 10^{-8}\; \Omega s$} & \multicolumn{1}{c|}{$\eta=0.01$} \\ \hline\hline
\multicolumn{4}{|c|}{Parameters found by fitting the rising output transition case} \\
\multicolumn{1}{|c|}{$R_p=2.912226\; k\Omega$} & \multicolumn{1}{c|}{$\alpha_3=0.157\cdot 10^{-8}\; \Omega s$} & \multicolumn{1}{c|}{$\alpha_4=0.161\cdot 10^{-8}\; \Omega s$} & \multicolumn{1}{c|}{$\eta=0.01$} \\ \hline
\end{tabular}}
\end{table}

\section{Conclusions}	
\label{sec:conclusions}

We provided a novel approach for developing hybrid delay models for dynamic
timing analysis of CMOS gates, and showed that it is capable of faithfully
covering all MIS effects in the case of a \NOR\ gate and a Muller \C\ gate.
\ifthenelse{\boolean{conference}}
{}
{
Relying on a first-order model of dimension 1, the resulting model for the \NOR\ gate is even simpler than 
the immediate switch hybrid ODE model proposed in \cite{FMOS22:DATE}, albeit
it does not share its failure to model the MIS effect for rising output transitions. 
Thanks to its simplicity, it allows to compute accurate approximation formulas 
for the gate delays, which makes the model efficiently applicable in dynamic digital 
timing analysis and facilitates easy model parametrization.}
An experimental 
comparison of the modeling accuracy against alternative approaches confirmed its superior performance.

\bibliographystyle{IEEEtran}
\bibliography{mybib}

\end{document}